\documentclass[journal]{IEEEtran}

\ifCLASSINFOpdf
\else
\fi

\usepackage{scalefnt}
\usepackage[shortlabels]{enumitem}
\usepackage{graphicx}
\usepackage{amssymb}
\usepackage{amsmath}
\usepackage{amsfonts}
\usepackage{color}
\usepackage{cite}
\usepackage{multicol}
\usepackage{theorem}
\usepackage{hyperref}
\usepackage{amsmath}
\usepackage{float}
\usepackage[caption=false]{subfig}
\usepackage{algorithm}
\usepackage{algorithmic}
\usepackage[normalem]{ulem}
\usepackage[dvipsnames]{xcolor}
\usepackage{multirow}

\begin{document}
\scalefont{0.861}%
\bstctlcite{IEEEexample:BSTcontrol}
\title{A Unified KLD Framework for Duplexity and Deployment Paradigms in Cell-Free mMIMO-ISAC}
\author{Yousef Kloob,~\IEEEmembership{Member,
~IEEE}, Mohammad Al-Jarrah,~\IEEEmembership{Member,
~IEEE}, and Emad Alsusa,~\IEEEmembership{Senior Member,~IEEE}\vspace{-0.4in}\thanks{%
Y. Kloob, M. Al-Jarrah, and E. Alsusa are with the Department of Electrical and
Electronic Engineering, University of Manchester, Manchester M13 9PL, U.K.
(e-mail: \{yousef.kloob, mohammad.al-jarrah, e.alsusa\}@manchester.ac.uk)}\thanks{%
}}
\maketitle

\begin{abstract}
This paper presents a comparative study of four potential operating configurations for distributed cell-free massive multiple-input multiple-output (CF-mMIMO) integrated sensing and communication (ISAC), spanning separated (SE) and shared (SH) access point (AP) deployment with half-duplex (HD) and full-duplex (FD) paradigms. The system comprises distributed APs serving multiple downlink (DL) and uplink (UL) users while simultaneously detecting radar targets. The configurations incorporate realistic impairments at the AP receivers: residual self-interference (SI) from transmit--receive leakage under FD operation, imperfect interference cancellation (IC) of the known radar and DL waveforms due to channel-estimation errors, and environmental clutter. To establish a common analytical scale for communication and sensing, the Kullback--Leibler divergence (KLD) is adopted as a unifying measure that represents both subsystems in comparable quantities, thereby enabling consistent comparison between error-rate and detection metrics. A generalised likelihood ratio test (GLRT) framework is developed, yielding closed-form expressions that link the KLD to the detection probability. Our results confirm the derived KLD-to-symbol error rate (SER) and KLD-to-detection links: with adequate SI suppression and IC quality, FD attains substantial communication gains over HD while preserving strong radar detection, and SH deployment raises both communication and radar performance through its larger effective aperture, though its radar gain then depends on cancellation quality, which SE deployment avoids by isolating the subsystems. These trends persist under imperfect channel state information (CSI) and sensing estimation, and a complexity analysis attributes the SH deployment and FD gains to a higher per-configuration processing cost, yielding deployment guidelines and quantitative design thresholds for next-generation CF-mMIMO ISAC systems.
\end{abstract}


\IEEEpeerreviewmaketitle

\vspace{-0.1 in}
\section{Introduction}
\IEEEPARstart{T}{he} rapid evolution of wireless communication networks has given rise to sensing-dependent applications, including autonomous driving, unmanned aerial vehicles (UAVs), and Internet-of-Things (IoT) devices \cite{9779322,9509294}. As the sixth-generation (6G) era approaches, network operators aim to extend their services beyond conventional communication to encompass sensing functionalities such as detection, localisation, tracking, and environmental surveillance \cite{10536135,9861699}. This dual requirement has positioned integrated sensing and communication (ISAC) systems as a cornerstone technology for next-generation wireless networks, where base station resources are synergistically utilised for both purposes \cite{9705498,9226446}.

Likewise, cell-free massive multiple-input multiple-output (CF-mMIMO) architectures have emerged as a promising paradigm for 6G networks, wherein distributed access points (APs) coherently serve user equipments (UEs) without traditional cell boundaries \cite{7827017}. By distributing antennas across a wide geographical area, CF-mMIMO systems achieve macro-diversity gains that enhance coverage uniformity and mitigate inter-cell interference \cite{10684238,9064545}, offering inherent advantages for simultaneous multi-point transmission and reception.

The duplexity paradigm, whether half-duplex (HD) or full-duplex (FD), fundamentally influences spectral efficiency and operational complexity. In HD operation, transmission and reception occur in orthogonal channels, e.g., separate time slots or different frequency bands, thereby avoiding self-interference (SI) at the cost of reduced resource utilisation. Conversely, FD enables simultaneous in-band transmission and reception, potentially doubling spectral efficiency, but introduces SI that must be suppressed through interference cancellation (IC) techniques \cite{10463523,10158724}. The ISAC deployment scenario further compounds this complexity: in separated (SE) deployment, distinct antenna subsets are allocated to different subsystems, whereas in shared (SH) deployment, all antennas serve multiple functionalities simultaneously \cite{8288677,9200993}.

\vspace{-0.12 in}
\subsection{Literature Review}
\vspace{-0.02 in}
Evaluating ISAC systems has traditionally relied on disparate metrics for the communication and sensing subsystems. Communication performance is typically characterised through achievable rate, outage probability, and bit error rate (BER), whereas sensing performance is assessed via detection probability, false alarm probability, and mean square error \cite{9800940}. This disparity complicates holistic assessment and joint optimisation. To address this, unified performance measures have been introduced, with mutual information-based approaches explored as a common framework \cite{10.5555/1146355,7086341,4776572}. Mutual information quantifies the average information shared between random variables and is therefore naturally connected to rate-oriented analysis, though such formulations typically assume capacity-achieving codes. More recently, the Kullback--Leibler divergence (KLD), also known as relative entropy, has emerged as a unified distinguishability measure for ISAC system design \cite{kloob,Al-Jarrah2023,kloob1,10118838}. Unlike mutual information, KLD measures the separation between specific probability distributions, and thus it provides a direct decision-level interpretation for finite-length symbol detection and radar hypothesis testing through Stein's lemma.
\label{r12a}

The application of KLD in sensing systems is well-established \cite{7086341,5467189}, particularly for MIMO radar waveform design in the presence of clutter and interference. However, its potential for characterising communication performance has only recently been recognised. In our previous works \cite{Al-Jarrah2023,10118838}, we have employed KLD to analyse the performance trade-off in ISAC systems with SE antenna deployment, establishing analytical relationships linking the detection probability $P_\mathrm{D}$ and the communication BER to the achievable KLD. This connection arises from the fundamental relationship between KLD and maximum-likelihood (ML) detection, the cornerstone of both radar and communication receivers. In \cite{kloob}, a low-complexity unified objective function is proposed based on KLD for optimising network resources in both SE and SH deployment scenarios, while the achievable KLD trade-off in multi-user multi-target (MUMT) ISAC systems is investigated in \cite{Fan0,11288098}. These efforts demonstrate that KLD provides a principled and analytically tractable framework for ISAC system design, enabling direct comparison and joint optimisation on a common scale. { However, these prior KLD-based works consider a co-located MIMO base station serving downlink (DL) users under HD operation.\label{r12g}}

The application of FD-based ISAC has been investigated in \cite{9724187,10159012}, where joint transmit and receive beamforming designs are proposed to manage the simultaneous presence of communication and radar signals, demonstrating that FD-ISAC can achieve significant spectral efficiency improvements when SI is adequately suppressed. Nevertheless, existing FD-ISAC studies primarily focus on co-located architectures and do not address the unique characteristics of distributed CF-mMIMO deployments, where both intra-AP SI and inter-AP mutual interference must be jointly managed. On the other hand, the integration of CF-mMIMO architectures into ISAC applications has been explored in \cite{10494224,10742291}, where the distributed topology enhances both communication coverage and sensing accuracy compared to co-located deployments. The macro-diversity inherent in CF-mMIMO translates to improved target detection and reduced outage probability for communication users. However, existing CF-mMIMO ISAC studies predominantly consider HD operation and compare neither the duplexity paradigms nor the deployment scenarios.

Despite these advances, several critical gaps remain. First, there is a lack of comprehensive analytical frameworks that simultaneously address both duplexity paradigms (HD and FD) and both deployment scenarios (SE and SH) within a unified CF-mMIMO ISAC context. Second, existing works do not adequately characterise the practical impairments, such as residual SI, imperfect IC, and environmental clutter, that influence the performance trade-offs between HD and FD operation. Third, the literature lacks actionable deployment guidelines specifying the SI suppression levels and IC quality requirements under which FD becomes advantageous over HD in distributed ISAC systems.

\vspace{-0.11 in}
\subsection{Motivation and Contributions}
\vspace{-0.02 in}
Motivated by the aforementioned gaps, this paper presents a comprehensive framework for analysing duplexity paradigms in distributed CF-mMIMO ISAC systems. Our framework encompasses four operational configurations: SE deployment HD (SE-HD), SE deployment FD (SE-FD), SH deployment HD (SH-HD), and SH deployment FD (SH-FD), each incorporating realistic impairments essential for practical system designs. The system model comprises multiple distributed APs serving DL and uplink (UL) UEs while simultaneously detecting multiple radar targets, with zero-forcing (ZF) beamforming for communication and identity covariance design for radar. In contrast to our published works in \cite{kloob,kloob1, Al-Jarrah2023,10118838,11288098,11036875}, which have considered a single MIMO base station with HD operation, the present paper applies KLD to a distributed CF-mMIMO architecture and, in terms of scope, provides a comparative performance analysis that jointly spans the HD and FD paradigms and the SE and SH deployment scenarios. Consequently, this work inherently analyses UL and DL communication and investigates the system robustness under practical imperfections, including the residual SI arising due to FD operation, and imperfect channel state information (CSI) and sensing estimation. To this end, the key contributions are summarised as follows:

\begin{itemize}[leftmargin=1em]

\item We provide a comparative analysis across the four configurations, spanning the SE and SH deployment scenarios and the HD and FD paradigms. SH deployment with a larger effective aperture raises communication and radar performance, though its radar gain then depends on cancellation quality, which SE deployment avoids by isolating the subsystems. FD is shown to improve communication over HD when the SI suppression and IC quality are adequate, while retaining strong radar detection. These trends remain robust under imperfect CSI and sensing estimation, with the deployment and duplexity ordering preserved.

\item We establish a unified system and signal model spanning the four configurations, in which the duplexity- and deployment-dependent residual SI, imperfect IC, and clutter are collected into a common set of interference-plus-noise variances, providing the consistent basis on which the four configurations are compared.

\item Utilising the KLD measure, we place communication and radar on a common scale, overcoming the limitation of disparate performance measures. Closed-form communication KLD expressions are derived for DL and UL and linked to the symbol error rate (SER), while the radar KLD is linked to detection performance through the generalised likelihood ratio test (GLRT) framework and the miss-detection exponent of Stein's lemma; both relations are derived and corroborated.

\item A GLRT-based detection framework is developed for the radar subsystem, with closed-form expressions for the test statistic distribution under both hypotheses. The framework provides ML estimates of unknown target parameters while maintaining analytical tractability, with the detection threshold calculated to satisfy a prespecified false alarm probability.

\item We analyse the per-configuration computational complexity of the four configurations, quantifying the cost of ZF precoding and combining, GLRT detection, and FD SI cancellation. The gains of SH deployment and FD operation are shown to come at a higher processing cost, adding an implementation-cost dimension to the deployment guidelines.

\end{itemize}
\label{r32}

The rest of this paper is organised as follows. Section~II presents the system model for all four operational configurations, including the signal models and impairment characterisation. Section~III develops the GLRT-based radar detection framework and the associated detection-probability metrics. Section~IV derives the unified KLD analysis for the communication and radar subsystems, covering DL, UL, and sensing, and establishes its links to SER and detection probability. Section~V presents the numerical results and discusses the performance trade-offs. Finally, Section~VI concludes the paper.

\textit{Notation:} Bold uppercase and lowercase letters denote matrices and vectors, respectively. Superscripts ${(\cdot)}^T$ and ${(\cdot)}^H$ denote transpose and Hermitian transpose. Subscripts ${(\cdot)}_\mathrm{c}$ and ${(\cdot)}_\mathrm{r}$ relate to communication and radar subsystems. The operators $\|\cdot\|$, $\|\cdot\|_F$, $\operatorname{tr}(\cdot)$, and $|\cdot|$ denote Euclidean norm, Frobenius norm, trace, and absolute value. The symbol $\odot$ represents the Hadamard product, and $\mathbf{I}_N$ is the $N \times N$ identity matrix. 
\vspace{-0.07 in}

\section{System Model}

\begin{table}[t]
\centering
\caption{Summary of the main system-model notation.}
\vspace{-0.05 in}
\label{tab:notation}
\renewcommand{\arraystretch}{1.12}\footnotesize\setlength{\tabcolsep}{4pt}
\begin{tabular}{@{}l p{0.66\columnwidth}@{}}
\hline\hline
\multicolumn{2}{@{}l}{\textit{Conventions}}\\
\hline
$(\cdot)^{(\varsigma)}$ & Configuration-dependent quantity, $\varsigma\in\mathcal{S}$\\
$\dot{(\cdot)},\grave{(\cdot)},\hat{(\cdot)}$ & SH deployment, UL, estimated quantity\\
\hline
\multicolumn{2}{@{}l}{\textit{Dimensions and configurations}}\\
\hline
$\varsigma,\mathcal{S}$ & Operating configuration and configuration set\\
$M,M_{\mathrm{r}},M_{\mathrm{c}}$ & Total / radar / communication APs (SE)\\
$M^{(\varsigma)}$ & Radar aperture, $M^{(\varsigma)}\in\{M_{\mathrm{r}},M\}$\\
$K_{\mathrm{D}},K_{\mathrm{U}}$ & DL / UL UEs\\
$T,L$ & Radar targets (bins) / snapshots per frame\\
$\beta,\beta_{\mathrm{AP}},\beta_{\mathrm{R}}$ & Transmit--receive SI leakage coefficients\\
\hline
\multicolumn{2}{@{}l}{\textit{Power}}\\
\hline
$P_{\mathrm{c}},P_{\mathrm{u}},P_{\mathrm{r}}$ & DL / UL / radar power ($P_{\mathrm{c},k},\grave{P}_{\mathrm{c},k},P_{\mathrm{r},t}$ per UE / radar beam)\\
$\mathbf{P}_{\mathrm{c}},\grave{\mathbf{P}}_{\mathrm{c}},\mathbf{P}_{\mathrm{r}}$ & DL / UL / radar power-allocation matrices\\
$\mathbf{R}_{x},\dot{\mathbf{R}}_{x}$ & Radar waveform covariance, SE / SH\\
\hline
\multicolumn{2}{@{}l}{\textit{Channels, large-scale fading, and target geometry}}\\
\hline
$\mathbf{H},\grave{\mathbf{H}}$ & SE DL AP-to-DL-UE / SE UL UE-to-communication-AP channels\\
$\dot{\mathbf{H}},\dot{\mathbf{F}}$ & SH DL AP-to-DL-UE channel / clutter channel\\
$\mathbf{F}$ & Radar-AP-to-DL-UE interference channel (SE)\\
$\mathbf{H}_{\mathrm{u}}$ & UE-to-UE interference channel (FD)\\
$\mathbf{H}_{\mathrm{SI}},\dot{\mathbf{H}}_{\mathrm{SI}}$ & Effective SI channel, SE and SH\\
$\mathbf{H}_{\mathrm{R2C}}$ & Radar-to-communication backscattering\\
$\mathbf{H}_{\mathrm{R}},\mathbf{H}_{\mathrm{R2I}}$ & Radar-to-communication / radar SI channel (SE, FD)\\
$\mathbf{G}_{\mathrm{r}},\dot{\mathbf{G}}_{\mathrm{r}}$ & Communication-to-radar interference / clutter\\
$\mathbf{G}_{\mathrm{r}}^{(\mathrm{UL})},\dot{\mathbf{G}}_{\mathrm{r}}^{(\mathrm{UL})}$ & UL-UE-to-radar-AP channel, SE / SH\\
$\mathbf{D}_{\mathrm{c}},\grave{\mathbf{D}}_{\mathrm{c}},\mathbf{D}_{\mathrm{u}}$ & DL AP-to-UE / UL UE-to-AP / UE-to-UE large-scale fading\\
$\mathbf{D}_{\mathrm{ru}},\grave{\mathbf{D}}_{\mathrm{ru}},\mathbf{D}_{\mathrm{rc}}$ & Radar-AP-to-DL-UE / UL-UE-to-radar-AP / radar-to-communication fading\\
$\mathbf{D}_{\mathrm{r},t}$ & Rank-one two-way target pathloss\\
$\mathbf{a}_t,\mathbf{A}_t,\boldsymbol{\Theta}_t,\mathbf{g}_t^{(\varsigma)}$ & Steering vector, response $\mathbf{a}_t\mathbf{a}_t^T$, phase alignment, effective $\mathbf{d}_{\mathrm{r},t}{\odot}\boldsymbol{\Theta}_t\mathbf{a}_t$\\
$\alpha_t$ & Complex target reflection coefficient\\
\hline
\multicolumn{2}{@{}l}{\textit{Interference residuals and variances}}\\
\hline
$\omega_{\mathrm{d},k}^{(\varsigma)}$ & Aggregate DL interference-plus-noise; var. $\sigma_{\omega,\mathrm{d},k}^{2(\varsigma)}$\\
$\boldsymbol{\omega}_{\mathrm{c}}^{(\varsigma)}$ & Post-IC UL residual; var. $\sigma_{\omega,\mathrm{c}}^{2(\varsigma)}$\\
$\boldsymbol{\omega}_{\mathrm{r}}^{(\varsigma)}$ & Post-IC radar residual; var. $\sigma_{\omega,\mathrm{r}}^{2(\varsigma)}$\\
$\sigma_{\xi,k}^{2(\varsigma)}$ & UL variance after ZF combining\\
$\sigma_{n}^{2},\sigma_{H}^{2},\sigma_{F}^{2},\sigma_{\mathrm{u}}^{2}$ & Noise / channel / clutter / UE-to-UE variance\\
$\sigma_{\mathrm{w}}^{2},\sigma_{\dot{\mathrm{w}}}^{2}$ & SE / SH precoder entry variance\\
$\sigma_{\{\mathrm{U,R}\}\{\mathrm{H,F}\},\mathrm{err}_i}^{2}$ & UL/radar (U/R), HD/FD (H/F) IC-error var.; $i{=}1$ SE, $2$ SH\\
\hline\hline
\end{tabular}
\vspace{-0.2 in}
\end{table}

This paper considers a MUMT distributed CF-mMIMO scenario. The ISAC system comprises $M$ single-antenna APs and $K = K_{\mathrm{D}} + K_{\mathrm{U}}$ single-antenna UEs, where $K_{\mathrm{D}}$ UEs operate in the DL and $K_{\mathrm{U}}$ UEs operate in the UL, while simultaneously detecting up to $T$ radar targets. In HD operation, each frame consists of two phases: (i)~a DL/radar-transmission phase where APs transmit to DL UEs and emit radar waveforms, and (ii)~a UL/echo-reception phase where APs receive UL signals and collect target echoes; the frame-level timing alignment this requires is substantiated in Section~\ref{subsec:sep_deploy}. In FD operation, the APs simultaneously transmit and receive, while UEs remain HD. For FD, we assume in-band simultaneous transmit and receive (STAR) operation on a shared monostatic aperture, where the same element supports concurrent transmission and reception over the same time--frequency resource \cite{10463523,10158724}, producing residual transmit--receive leakage proportional to the transmitted waveform, captured through a small leakage coefficient $\beta$ scaling the effective SI channels; in HD, no such leakage exists. The considered setup can also be interpreted as one coordinated region of a larger clustered CF-mMIMO deployment, so the processing is applied within the coordinated region rather than over an arbitrarily large network \cite{9064545}.

The central processing unit (CPU) coordinates a network-wide power budget \(P_\mathrm{T}\), allocated to DL communication, UL communication \cite{7827017}, and radar subsystems as $P_\mathrm{c}$, $P_\mathrm{u}$, and $P_\mathrm{r}$, respectively, with \(P_\mathrm{T}=P_\mathrm{c}+P_\mathrm{u}+P_\mathrm{r}\), \(P_\mathrm{c}=\sum_{k=1}^{K_{\mathrm D}}P_{\mathrm c,k}\), \(P_\mathrm{r}=\sum_{t=1}^{T}P_{\mathrm r,t}\), and \(P_\mathrm{u}=\sum_{k=1}^{K_{\mathrm U}}\grave P_{\mathrm c,k}\)~\cite{10494224}. ZF beamforming is employed to precode the DL UE information while eliminating inter-user interference \cite{Fatema2018MassiveML}. The radar waveform covariance matrix is $\mathbf{R}_{x}\triangleq \frac{1}{L}\sum_{l=1}^{L}\mathbf{x}_{l}\mathbf{x}_{l}^{H}\in \mathbb{C}^{M_\mathrm{r}\times M_\mathrm{r}}$, where $L$ is the number of snapshots and $\mathbf{x}_l\in \mathbb{C}^{M_\mathrm{r} \times 1}$ is the power-normalised radar waveform vector for snapshot $l$. We also define the index sets \(\mathcal K_{\mathrm D}\triangleq\{1,\ldots,K_{\mathrm D}\}\), \(\mathcal K_{\mathrm U}\triangleq\{1,\ldots,K_{\mathrm U}\}\), \(\mathcal T\triangleq\{1,\ldots,T\}\), and \(\mathcal L\triangleq\{1,\ldots,L\}\) for the DL UEs, UL UEs, radar targets, and snapshots, respectively.

 Let $\varsigma\!\in\!\mathcal{S}\!\triangleq\!\{\mathrm{SE\mbox{-}HD},\mathrm{SE\mbox{-}FD},\mathrm{SH\mbox{-}HD},\mathrm{SH\mbox{-}FD}\}$ denote the operating configuration; a quantity carrying the superscript $(\varsigma)$ is configuration-dependent, and the configuration-dependent interference and residual terms are collected in the case-defined expressions of this section. A dot marks SH deployment quantities, namely channels, precoders, and received signals, and its absence marks SE deployment, while the grave and hat accents denote UL and estimated quantities; in raw form the received signals additionally carry a superscript giving the link direction and duplexity, such as $\mathbf{y}_l^{(\mathrm{DL,HD})}$ and $\dot{\mathbf{y}}_l^{(\mathrm{UL,FD})}$, and the post-IC quantities carry the configuration superscript $(\varsigma)$. These conventions, summarised in Table~\ref{tab:notation}, are used throughout Sections~III and~IV.

\vspace{-0.1in}
\subsection{Separated Deployment}\label{subsec:sep_deploy} 
In SE deployment, the APs are divided into two disjoint subsets: $M_\mathrm{r}$ APs dedicated to radar operation, and $M_\mathrm{c}=M-M_\mathrm{r}$ APs for communication. Under HD operation, time-division duplex (TDD) is adopted, so that each snapshot interval $l$ comprises two phases: during DL, communication APs transmit to DL UEs while radar APs emit radar signals; during UL, communication APs receive UL signals and radar APs capture target echoes. This TDD structure assumes frame-level timing alignment among the APs. This assumption is consistent with TDD distributed multi-user MIMO, where scalable AP synchronisation has been shown to provide the accuracy required for coherent joint precoding~\cite{6760595}. Under FD operation, the communication APs perform DL transmission and UL reception concurrently, while the radar APs continuously transmit and receive within each snapshot $l$. This concurrency introduces residual transmit--receive leakage in both subsystems, modelled by the coupling coefficients $\beta_{\mathrm{AP}}$ and $\beta_{\mathrm{R}}$ (with $0\le\beta_{\mathrm{AP}},\beta_{\mathrm{R}}\ll1$) that scale the residual SI channels~\cite{10463523,10158724}, treated as independent parameters since the two receive paths can exhibit different residual coupling levels. The UEs remain HD throughout, with DL UEs only receiving and UL UEs only transmitting. 

\subsubsection{Communication System: Downlink}
At each snapshot $l$, data symbols $s_{k,l}$ for the $k$-th DL UE are drawn from a normalised constellation with $\mathbb{E}\{ \left\vert s_{k,l} \right\vert ^{2}\} =1$. Given the DL channel matrix $\mathbf{H}\in \mathbb{C}^{K_{\mathrm{D}}\times M_{\mathrm{c}}}$ from the communication APs to the DL UEs, the symbols are precoded using a ZF precoder $\mathbf{W}_{\mathrm{c},l}=\alpha_{\mathrm{ZF}}{\tilde{\mathbf{W}}_{\mathrm{c},l}}$, where $\alpha_{\mathrm{ZF}}=\frac{1}{\sqrt{\mathbf{s}_{l}^{H}\tilde{\mathbf{W}}_{\mathrm{c},l}^{H}\tilde{\mathbf{W}}_{\mathrm{c},l}\mathbf{s}_{l}}}$ is the normalisation factor ensuring power constraint satisfaction, and $\tilde{\mathbf{W}}_{\mathrm{c},l}=\mathbf{H}^{H}\left(\mathbf{H}\mathbf{H}^{H}\right)^{-1}$ is the non-normalised ZF precoder. The channels $\mathbf{H}\in \mathbb{C}^{K_{\mathrm{D}}\times M_{\mathrm{c}}}$ and $\mathbf{F}\in \mathbb{C}^{K_{\mathrm{D}}\times M_{\mathrm{r}}}$ are modelled as flat Rayleigh fading with independent and identically distributed (i.i.d.)\ entries $\mathcal{CN}(0,2\sigma_H^2)$ and $\mathcal{CN}(0,2\sigma_F^2)$, respectively, where $\mathbf{F}$ is the interference channel from radar APs to DL UEs, and the receiver noise is $\mathbf{n}_{l}\sim\mathcal{CN}(\mathbf{0},2\sigma_{n}^{2}\mathbf{I}_{K_{\mathrm{D}}})$. Large-scale fading is captured by $\mathbf{D}_\mathrm{c}\in \mathbb{R}^{K_{\mathrm{D}}\times M_{\mathrm{c}}}$ and $\mathbf{D}_{\mathrm{ru}}\in \mathbb{R}^{K_{\mathrm{D}}\times M_{\mathrm{r}}}$ for the communication-AP-to-DL-UE and radar-AP-to-DL-UE links, respectively, with $[\mathbf{D}_\mathrm{c}]_{k,m}=d_{\mathrm{c},k,m}^{-\eta/2}$ and $[\mathbf{D}_{\mathrm{ru}}]_{k,n}=d_{\mathrm{ru},k,n}^{-\eta/2}$, where $d_{\mathrm{c},k,m}$ is the distance from communication AP $m$ to DL UE $k$, $d_{\mathrm{ru},k,n}$ the distance from radar AP $n$ to DL UE $k$, and $\eta$ is the pathloss exponent. With $\mathbf{P}_\mathrm{c}=\mathrm{diag}(\sqrt{P_{\mathrm{c},1}},\cdots,\sqrt{P_{\mathrm{c},K_{\mathrm{D}}}})$ the DL power-control matrix, the HD DL received signal $\mathbf{y}_{l}^{(\mathrm{DL,HD})} \in \mathbb{C}^{K_{\mathrm{D}} \times 1}$ is \cite{9064545},
    \vspace{-0.05 in}
\begin{equation}
\mathbf{y}_{l}^{(\mathrm{DL,HD})}
=
(\mathbf{D}_\mathrm{c}\!\odot\!\mathbf{H})
\mathbf{W}_{\mathrm{c},l}\mathbf{P}_\mathrm{c}\mathbf{s}_{l}
+
\sqrt{\frac{P_{\mathrm{r}}}{M_{\mathrm{r}}}}
(\mathbf{D}_{\mathrm{ru}}\!\odot\!\mathbf{F})\mathbf{x}_{l}
+\mathbf{n}_l,
\label{eq:1}
\vspace{-0.02 in}
\end{equation}
where the second term is the radar-induced interference. Under FD, the DL UEs receive while the UL UEs transmit in the same time--frequency resource, so the received signal additionally contains inter-UE interference,
\vspace{-0.04 in}
\begin{align}
\mathbf{y}_{l}^{(\mathrm{DL,FD})}
&=
(\mathbf{D}_{\mathrm{c}}\!\odot\!\mathbf{H})
\mathbf{W}_{\mathrm{c},l}\mathbf{P}_{\mathrm{c}}\mathbf{s}_{l}
+
\sqrt{\frac{P_{\mathrm{r}}}{M_{\mathrm{r}}}}
(\mathbf{D}_{\mathrm{ru}}\!\odot\!\mathbf{F})\mathbf{x}_{l}
\nonumber\\
&\quad+
(\mathbf{D}_{\mathrm{u}}\!\odot\!\mathbf{H}_{\mathrm{u}})
\grave{\mathbf{P}}_{\mathrm{c}}\mathbf{u}_{l}
+
\mathbf{n}_l,
\label{eq:DL_SEFD}
\vspace{-0.02 in}
\end{align}
where $\mathbf{D}_{\mathrm{u}} \in \mathbb{R}^{K_{\mathrm{D}} \times K_{\mathrm{U}}}$ captures the UL-UE-to-DL-UE pathloss with $[\mathbf{D}_{\mathrm{u}}]_{i,j} = d_{\mathrm{u},i,j}^{-\eta/2}$, $\forall i\in\mathcal{K}_{\mathrm{D}}$, $j\in\mathcal{K}_{\mathrm{U}}$, and $\mathbf{H}_{\mathrm{u}} \in \mathbb{C}^{K_{\mathrm{D}} \times K_{\mathrm{U}}}$ is the UE-to-UE interference channel with i.i.d. entries $\mathcal{CN}(0, 2\sigma_{\mathrm{u}}^{2})$. As the UEs are HD, UE SI is absent; the third term in \eqref{eq:DL_SEFD} is the only structural difference from \eqref{eq:1}. Let $\omega_{\mathrm{d},k}^{(\varsigma)}$ denote the aggregate interference-plus-noise at the $k$-th DL UE, comprising every term of \eqref{eq:1} under HD, or of \eqref{eq:DL_SEFD} under FD, other than the desired signal, that is, the radar-induced interference, the inter-UE interference (FD only), and the noise. These constituents arise from independent propagation channels, data symbols, and receiver noise, and are therefore mutually independent and zero-mean; moreover, the radar-induced term aggregates many independent contributions across the array, so $\omega_{\mathrm{d},k}^{(\varsigma)}$ is, by the central limit theorem (CLT), circularly-symmetric complex Gaussian, $\omega_{\mathrm{d},k}^{(\varsigma)}\sim \mathcal{CN}\bigl(0,2\sigma_{\omega,\mathrm{d},k}^{2(\varsigma)}\bigr)$ \cite{7827017,10494224}. This approximation, together with the analogous CLT-based models used for the UL and radar residuals below, holds in the regime $M\gg K_{\mathrm{D}}+K_{\mathrm{U}}+T$ characteristic of CF-mMIMO, in which the number of APs far exceeds the served users and sensed targets, and improves as $M$ grows. The per-dimension variance $\sigma_{\omega,\mathrm{d},k}^{2(\varsigma)}$ is
\vspace{-0.05 in}
\begin{equation}
\sigma_{\omega,\mathrm{d},k}^{2(\varsigma)}=
\begin{cases}
\frac{P_{\mathrm{r}}\operatorname{tr}(\mathbf{R}_{x})}{M_{\mathrm{r}}^{2}}\sum_{n=1}^{M_{\mathrm{r}}}d_{\mathrm{ru},k,n}^{-\eta}\sigma_{F}^{2}+\sigma_{n}^{2}, & \varsigma=\mathrm{SE\mbox{-}HD},\\[2mm]
\begin{aligned}[b]
&\dfrac{P_{\mathrm{r}}\operatorname{tr}(\mathbf{R}_{x})}{M_{\mathrm{r}}^{2}}\sum_{n=1}^{M_{\mathrm{r}}}d_{\mathrm{ru},k,n}^{-\eta}\sigma_{F}^{2}\\
&+\sum_{j=1}^{K_{\mathrm{U}}}\grave{P}_{\mathrm{c},j}d_{\mathrm{u},k,j}^{-\eta}\sigma_{\mathrm{u}}^{2}+\sigma_{n}^{2},
\end{aligned} & \varsigma=\mathrm{SE\mbox{-}FD}.
\end{cases}
\label{eq:SE_DL_var}
\vspace{-0.02 in}\end{equation}

\subsubsection{Communication System: Uplink}
During the UL phase, the signals received at the communication APs are forwarded to the CPU. Let $\mathbf{u}_l=[u_{1,l},\dots,u_{K_{\mathrm{U}},l}]^T\in\mathbb{C}^{K_{\mathrm{U}}\times 1}$ collect the UL data symbols with $\mathbb{E}\{|u_{k,l}|^2\}=1$, and $\grave{\mathbf{P}}_\mathrm{c}=\mathrm{diag}(\sqrt{\grave{P}_{\mathrm{c},1}},\dots,\sqrt{\grave{P}_{\mathrm{c},K_{\mathrm{U}}}})$ the UL transmit-power allocation. The UL channel matrix $\grave{\mathbf{H}}\in\mathbb{C}^{M_{\mathrm{c}}\times K_{\mathrm{U}}}$ has i.i.d. entries $\mathcal{CN}(0,2\sigma_H^2)$, with large-scale fading $\grave{\mathbf{D}}_\mathrm{c}\in\mathbb{R}^{M_{\mathrm{c}}\times K_{\mathrm{U}}}$, $[\grave{\mathbf{D}}_\mathrm{c}]_{m,k}=\grave{d}_{\mathrm{c},m,k}^{-\eta/2}$. The radar waveform transmitted by the radar APs is scattered by the environment and the targets and reaches the communication APs as an additional interference term; this radar-to-communication backscattered signal propagates through the channel $\mathbf{H}_{\mathrm{R2C}}\in\mathbb{C}^{M_{\mathrm{c}}\times M_{\mathrm{r}}}$, with i.i.d. entries $\mathcal{CN}(0,2\sigma_H^2)$ and pathloss $\mathbf{D}_{\mathrm{rc}}\in \mathbb{R}^{M_{\mathrm{c}}\times M_{\mathrm{r}}}$, $[\mathbf{D}_{\mathrm{rc}}]_{m,n}=d_{\mathrm{rc},m,n}^{-\eta/2}$. Under HD, the aggregate received signal is
    \vspace{-0.05 in}
\begin{equation}
\mathbf{y}_l^{(\mathrm{UL,HD})}
=
(\grave{\mathbf{D}}_\mathrm{c}\!\odot\!\grave{\mathbf{H}})
\grave{\mathbf{P}}_\mathrm{c}\mathbf{u}_{l}
+
\sqrt{\frac{P_{\mathrm{r}}}{M_{\mathrm{r}}}}
(\mathbf{D}_\mathrm{rc}\!\odot\!\mathbf{H}_{\mathrm{R2C}})
\mathbf{x}_{l}
+
\mathbf{n}_l,
\vspace{-0.02 in}
\end{equation}
whereas under FD the observation is additionally impaired by SI from the concurrent DL transmissions and by the radar interference,
\vspace{-0.04 in}
\begin{align}
\mathbf{y}_l^{(\mathrm{UL,FD})}
&=
(\grave{\mathbf{D}}_\mathrm{c}\!\odot\!\grave{\mathbf{H}})
\grave{\mathbf{P}}_\mathrm{c}\mathbf{u}_{l}
+
\beta_{\mathrm{AP}}\mathbf{H}_{\mathrm{SI}}
\mathbf{W}_{\mathrm{c},l}\mathbf{P}_\mathrm{c}\mathbf{s}_{l}
\nonumber\\
&\quad+
\sqrt{\frac{P_\mathrm{r}}{M_\mathrm{r}}}
(\mathbf{D}_\mathrm{rc}\!\odot\!\mathbf{H}_{\mathrm{R}})
\mathbf{x}_{l}
+
\mathbf{n}_l,
\label{eq:UL_SEFD_raw}
\vspace{-0.02 in}
\end{align}
where $\mathbf{H}_{\mathrm{SI}}\in\mathbb{C}^{M_{\mathrm{c}}\times M_{\mathrm{c}}}$ is the effective SI channel of the communication APs' DL transmissions (diagonal entries capture intra-AP leakage, off-diagonal entries inter-AP coupling), scaled by $\beta_{\mathrm{AP}}$, and $\mathbf{H}_{\mathrm{R}}\in\mathbb{C}^{M_{\mathrm{c}}\times M_{\mathrm{r}}}$ is the radar-to-communication interference channel with i.i.d. entries $\mathcal{CN}(0,2\sigma_H^2)$. Since the radar waveform $\mathbf{x}_l$, and under FD the DL waveform, are known at the CPU, the corresponding backscattered-signal and SI terms are suppressed via IC using the minimum mean-square error (MMSE) estimates $\hat{\mathbf{H}}_{\mathrm{R2C}}$, $\hat{\mathbf{H}}_{\mathrm{SI}}$, $\hat{\mathbf{H}}_{\mathrm{R}}$. After cancellation, the UL received signal for both duplexities takes the unified form
    \vspace{-0.05 in}
\begin{equation}
\mathbf{y}_{l}^{(\mathrm{UL},\varsigma)}
=
(\grave{\mathbf{D}}_{\mathrm{c}}\!\odot\!\grave{\mathbf{H}})
\grave{\mathbf{P}}_{\!\mathrm{c}}\mathbf{u}_{l}
+
\boldsymbol{\omega}_{\mathrm{c}}^{(\varsigma)}
+
\mathbf{n}_l,
\label{eq:UL_afterIC}
\vspace{-0.02 in}
\end{equation}
where the post-IC residual interference is
\vspace{-0.05 in}
\begin{equation}
\boldsymbol{\omega}_{\mathrm{c}}^{(\varsigma)}=
\begin{cases}
\sqrt{\tfrac{P_{\mathrm{r}}}{M_{\mathrm{r}}}}\bigl(\mathbf{D}_{\mathrm{rc}}\!\odot\!(\mathbf{H}_{\mathrm{R2C}}-\widehat{\mathbf{H}}_{\mathrm{R2C}})\bigr)\mathbf{x}_{l}, & \varsigma=\mathrm{SE\mbox{-}HD},\\[2mm]
\begin{aligned}[b]
&\beta_{\mathrm{AP}}\bigl(\mathbf{H}_{\mathrm{SI}}-\widehat{\mathbf{H}}_{\mathrm{SI}}\bigr)\mathbf{W}_{\mathrm{c},l}\mathbf{P}_{\mathrm{c}}\mathbf{s}_{l}\\
&+\sqrt{\tfrac{P_{\mathrm{r}}}{M_{\mathrm{r}}}}\bigl(\mathbf{D}_{\mathrm{rc}}\!\odot\!(\mathbf{H}_{\mathrm{R}}-\widehat{\mathbf{H}}_{\mathrm{R}})\bigr)\mathbf{x}_{l},
\end{aligned} & \varsigma=\mathrm{SE\mbox{-}FD}.
\end{cases}
\label{eq:SE_UL_residual}
\vspace{-0.02 in}\end{equation}
The HD residual stems from imperfect IC of the radar backscattered signal; the FD residual adds the imperfectly cancelled SI. Modelling the estimation errors in the HD and FD cases as i.i.d. $\mathcal{CN}(0,2\sigma_{\mathrm{UH},\mathrm{err}_1}^{2})$ and $\mathcal{CN}(0,2\sigma_{\mathrm{UF},\mathrm{err}_1}^{2})$, respectively, each entry of $\boldsymbol{\omega}_{\mathrm{c}}^{(\varsigma)}$ sums many independent zero-mean terms, so by the CLT, $\boldsymbol{\omega}_{\mathrm{c}}^{(\varsigma)}\sim\mathcal{CN}\bigl(\mathbf{0},\,2\sigma_{\omega,\mathrm{c}}^{2(\varsigma)}\mathbf{I}_{M_{\mathrm{c}}}\bigr)$ with
\vspace{-0.05 in}
\begin{equation}
\sigma_{\omega,\mathrm{c}}^{2(\varsigma)}=
\begin{cases}
\sigma_{\mathrm{UH},\mathrm{err}_1}^{2}\tfrac{P_{\mathrm{r}}}{M_{\mathrm c}M_{\mathrm{r}}^2}\|\mathbf{D}_{\mathrm{rc}}\|^2_F\operatorname{tr}(\mathbf{R}_{x}), & \varsigma=\mathrm{SE\mbox{-}HD},\\[2mm]
\begin{aligned}[b]
&\sigma_{\mathrm{UF},\mathrm{err}_1}^{2}\Bigl(\beta_{\mathrm{AP}}^2\sigma_{\mathrm{w}}^2 M_\mathrm{c}P_\mathrm{c}\\
&+\tfrac{P_{\mathrm{r}}}{M_{\mathrm c}M_{\mathrm{r}}^2}\|\mathbf{D}_{\mathrm{rc}}\|_{F}^{2}\operatorname{tr}(\mathbf{R}_{x})\Bigr),
\end{aligned} & \varsigma=\mathrm{SE\mbox{-}FD},
\end{cases}
\label{eq:SE_UL_var}
\vspace{-0.02 in}\end{equation}
where $\sigma_{\mathrm{w}}^{2}$ is the variance of the entries of the DL precoder $\mathbf{W}_{\mathrm{c},l}\triangleq[\mathbf{w}_{\mathrm{c},1,l},\ldots,\mathbf{w}_{\mathrm{c},K_{\mathrm{D}},l}]$, $\mathbf{w}_{\mathrm{c},k,l}\in\mathbb{C}^{M_{\mathrm{c}}\times 1}$, and $\sigma_{\dot{\mathrm{w}}}^{2}$ the corresponding entry variance of the SH deployment precoder $\dot{\mathbf{W}}_{\mathrm{c},l}\in\mathbb{C}^{M\times K_{\mathrm{D}}}$. The CPU then applies the linear ZF combiner $\mathbf{G}_{\mathrm{c}}=\bigl(\grave{\mathbf{H}}^{H}\grave{\mathbf{H}}\bigr)^{-1}\grave{\mathbf{H}}^{H}$ to $\mathbf{y}_{l}^{(\mathrm{UL},\varsigma)}$, yielding $\widetilde{\mathbf{u}}_l=\mathbf{G}_{\mathrm{c}}\mathbf{y}_l^{(\mathrm{UL},\varsigma)}\approx[\hat{u}_{1,l},\ldots,\hat{u}_{K_{\mathrm{U}},l}]^T$ as the estimate of the UL symbols.

\subsubsection{Radar System}
MIMO radar forms $T$ concurrent beams via orthogonal waveforms \cite{7126203, 6957532, 5728938, 9034082}, where $T$ is the number of probed radar bins per frame. The system assumes spatially distinct targets, each occupying a separate bin, with $T$ bounded by the available spatial degrees of freedom. Accordingly, the radar waveform formed at the CPU during the DL (radar transmission) phase of snapshot $l$ is
\vspace{-0.05 in}
\begin{equation}
\mathbf{x}_{l}=\mathbf{W}_{\mathrm{r},l}\:\mathbf{P}_\mathrm{r}\:\boldsymbol{\phi}_{l}, \label{eq:2}
\vspace{-0.02 in}\end{equation}
where $\boldsymbol{\phi}_{l}=\left[ \phi _{1,l},\phi _{2,l},\cdots ,\phi _{T,l}\right] ^{T}$ collects the snapshot-$l$ samples of $T$ orthonormal baseband waveforms satisfying $\frac{1}{L}\sum_{l=1}^{L}\boldsymbol{\phi}_{l}\boldsymbol{\phi}_{l}^{H}=\mathbf{I}_{T}$ \cite{9034082}, $\mathbf{P}_\mathrm{r}=\mathrm{diag}\!\left(\sqrt{P_{\mathrm{r},1}/{P_\mathrm{r}}},\cdots,\sqrt{P_{\mathrm{r},T}/{P_\mathrm{r}}}\right)$ is the dimensionless radar power-allocation matrix carrying the per-beam power fractions, so that the transmitted radar signal is $\sqrt{P_\mathrm{r}/M_\mathrm{r}}\,\mathbf{x}_{l}=\sum_{t=1}^{T}\sqrt{P_{\mathrm{r},t}/M_\mathrm{r}}\,\mathbf{w}_{\mathrm{r},t,l}\phi_{t,l}$, and $\mathbf{W}_{\mathrm{r},l}\in \mathbb{C}^{M_{\mathrm{r}}\times T}=[\mathbf{w}_{\mathrm{r},1,l},\cdots,\mathbf{w}_{\mathrm{r},T,l}]$ is the radar precoding matrix, with $\mathbf{w}_{\mathrm{r},t,l} \in \mathbb{C}^{M_{\mathrm{r}}\times 1}$ the precoding vector of the $t$-th beam. The radar precoder can be tailored to enhance sensing or to satisfy a desired transmit covariance constraint; for instance, an omnidirectional transmission is obtained by enforcing ${\mathbf{R}}_{\mathrm{w}} \triangleq \frac{1}{L}\sum_{l=1}^{L}\mathbf{W}_{\mathrm{r},l}\mathbf{W}_{\mathrm{r},l}^H=\mathbf{I}_{M_\mathrm{r}}$, where ${\mathbf{R}}_{\mathrm{w}}$ is the radar transmit covariance.

During echo reception, the radar APs forward the radar return signals to the CPU, where the returns are processed by a bank of matched filters tuned to $\phi_{t}$, $t\in\mathcal{T}$, each corresponding to a spatial--range--Doppler bin. Since $\phi_{t}\perp \phi_{i}$ for $t\neq i$, the per-bin detection is performed independently. For each bin $t$, the sensing geometry (and thus $\{d_{\mathrm{r},t,m}\}_{m=1}^{M_{\mathrm{r}}}$, $\mathbf{D}_{\mathrm{r},t}$, and $\mathbf{A}_t$) is fixed by the known AP locations and the bin parametrisation, while the complex reflection coefficient $\alpha_t$ is unknown and estimated via ML. We consider the binary hypothesis test $\mathcal{H}_{q}$, $q\in\{0,1\}$, where $q=0$ and $q=1$ denote target absence and presence. Under HD, the radar return signal for beam $t$ under $\mathcal{H}_{q}$ is
\vspace{-0.04 in}
\begin{align}
\mathbf{y}_{\mathrm{r},t,l|\mathcal{H}_{q}}^{(\mathrm{HD})}
=&\alpha_t\sqrt{\frac{
P_{\mathrm{r},t}}{M_{\mathrm{r}}}}
(\mathbf{D}_{\mathrm{r},t}\!\odot\!\mathbf{A}_t)
\mathbf{w}_{\mathrm{r},t,l}q
\nonumber\\
&+
(\grave{\mathbf{D}}_{\mathrm{ru}}\!\odot\!\mathbf{G}_\mathrm{r}^{(\mathrm{UL})})
\grave{\mathbf{P}}_\mathrm{c}\mathbf{u}_{l}
+
\mathbf{n}_{\mathrm{r},l},
\label{eq:3}
\vspace{-0.02 in}
\end{align}
where $\grave{\mathbf{D}}_{\mathrm{ru}}\in \mathbb{R}^{M_{\mathrm{r}}\times K_{\mathrm{U}}}$ captures the UL-UE--radar-AP pathloss, $[\grave{\mathbf{D}}_{\mathrm{ru}}]_{n,k}=\grave{d}_{\mathrm{ru},n,k}^{-\eta/2}$, and $\mathbf{G}_\mathrm{r}^{(\mathrm{UL})} \in \mathbb{C}^{M_{\mathrm{r}} \times K_{\mathrm{U}}}$ is the channel from the $K_{\mathrm{U}}$ UEs to the $M_{\mathrm{r}}$ radar APs; the second term is the UL-induced interference. The target pathloss is the vector $\mathbf{d}_{\mathrm{r},t}\triangleq\bigl[d_{\mathrm{r},t,1}^{-\eta/2},\ldots,d_{\mathrm{r},t,M_{\mathrm{r}}}^{-\eta/2}\bigr]^T$ with rank-one matrix $\mathbf{D}_{\mathrm{r},t}\triangleq \mathbf{d}_{\mathrm{r},t}\mathbf{d}_{\mathrm{r},t}^{T}\in\mathbb{R}^{M_{\mathrm{r}}\times M_{\mathrm{r}}}$, where $d_{\mathrm{r},t,m}$ is the one-way distance between the $m$-th radar AP and bin $t$, so that $[\mathbf{D}_{\mathrm{r},t}]_{m,m'}=(d_{\mathrm{r},t,m}\,d_{\mathrm{r},t,m'})^{-\eta/2}$ is the two-way pathloss of the path from AP $m'$ via the target to AP $m$, reducing to the monostatic round-trip pathloss $d_{\mathrm{r},t,m}^{-\eta}$ for $m'=m$. The scalar $\alpha_t$ models an isotropic point scatterer, i.e., a common bistatic response across the AP--target--AP paths, yielding the rank-one structure exploited by the unified GLRT and KLD derivations \cite{7089157}, and the noise is $\mathbf{n}_{\mathrm{r},l}\sim\mathcal{CN}(\mathbf{0},2\sigma _{n}^{2}\mathbf{I}_{M_\mathrm{r}})$. Defining the wavelength-normalised one-way distance $\tilde{d}_{\mathrm{r},t,m}\triangleq d_{\mathrm{r},t,m}/\lambda_0$, with $\lambda_0$ the wavelength, the distributed steering vector is
\begin{equation}
\mathbf{a}_{t}\triangleq \bigl[\mathrm{e}^{-j2\pi\tilde{d}_{\mathrm{r},t,1}},\,\mathrm{e}^{-j2\pi\tilde{d}_{\mathrm{r},t,2}},\,\cdots,\,\mathrm{e}^{-j2\pi\tilde{d}_{\mathrm{r},t,M_{\mathrm{r}}}}\bigr]^{T},
\label{eq:dist_steering}
\end{equation}
and the equivalent array response is $\mathbf{A}_t\triangleq \mathbf{a}_{t}\mathbf{a}_{t}^{T}\in \mathbb{C}^{M_{\mathrm{r}}\times M_{\mathrm{r}}}$, whose $(m,m')$-th entry $\mathrm{e}^{-j2\pi(\tilde{d}_{\mathrm{r},t,m}+\tilde{d}_{\mathrm{r},t,m'})}$ carries the accumulated two-way phase of the path from AP $m'$ via the target to AP $m$, with the monostatic diagonal $\mathrm{e}^{-j4\pi\tilde{d}_{\mathrm{r},t,m}}$. The APs are assumed phase-synchronised, which is attainable through over-the-air calibration in distributed MIMO \cite{6760595}, so the CPU can exploit these deterministic path phases through coherent multistatic processing \cite{godrich2010}; the phase common to all paths is absorbed into $\alpha_t$. Under FD, the radar APs transmit and receive concurrently within snapshot $l$, so that, beyond the target return and UL interference, the radar return signal contains communication-to-radar leakage and radar transmit--receive leakage,
\vspace{-0.04 in}
\begin{align}
\!\!\mathbf{y}_{\mathrm{r},t,l|\mathcal{H}_{q}}^{(\mathrm{FD})}
&=
\alpha_t
\sqrt{\!\frac{P_{\mathrm{r},t}}{M_{\mathrm{r}}}}
(\mathbf{D}_{\mathrm{r},t}\!\!\odot\!\!\mathbf{A}_t)
\mathbf{w}_{\mathrm{r},t,l}q
+
(\grave{\mathbf{D}}_{\mathrm{ru}}\!\odot\!\mathbf{G}_\mathrm{r}^{(\mathrm{UL})})
\grave{\mathbf{P}}_\mathrm{\!c}\mathbf{u}_l
\nonumber\\
&\quad+
(\mathbf{D}_\mathrm{rc}^{T}\!\odot\!\mathbf{G}_\mathrm{r})
\mathbf{W}_{\mathrm{c},l}\mathbf{P}_\mathrm{c}\mathbf{s}_l
+
\beta_{\mathrm{R}}\sqrt{\tfrac{P_{\mathrm{r}}}{M_{\mathrm{r}}}}
\mathbf{H}_{\mathrm{R2I}}\mathbf{x}_l
+
\mathbf{n}_{\mathrm{r},l},
\label{eq:SEFD_radar_raw}
\vspace{-0.02 in}
\end{align}
where $\mathbf{G}_\mathrm{r} \in \mathbb{C}^{M_\mathrm{r} \times M_\mathrm{c}}$ is the communication-to-radar DL interference channel and $\mathbf{H}_{\mathrm{R2I}} \in \mathbb{C}^{M_\mathrm{r} \times M_\mathrm{r}}$ the radar SI channel, both with i.i.d. entries $\mathcal{CN}(0,2\sigma_H^2)$. Since the CPU generates the DL and radar waveforms and recovers $\hat{\mathbf{u}}_l$ after UL decoding (UL channels estimated via TDD reciprocity), the terms cancelled by IC are the UL-induced interference in the HD case, together with the additional leakage terms in the FD case, so that the post-IC radar return signal for both duplexities takes the unified form
\vspace{-0.05 in}\begin{equation}
\mathbf{y}_{\mathrm{r},t,l|\mathcal{H}_{q}}^{(\varsigma)} =\alpha_t\sqrt{\frac{
P_{\mathrm{r},t}}{M_{\mathrm{r}}}}(\mathbf{D}_{\mathrm{r},t}\!\odot\!\mathbf{A}_t) 
\mathbf{w}_{\mathrm{r},t,l}q+\boldsymbol{\omega}_{\mathrm{r}}^{(\varsigma)},
\label{eq:SE_radar_afterIC}
\vspace{-0.02 in}\end{equation}
where the residual interference-plus-noise is
\vspace{-0.05 in}
\begin{equation}
\boldsymbol{\omega}_{\mathrm{r}}^{(\varsigma)}=
\begin{cases}
(\grave{\mathbf{D}}_{\mathrm{ru}}\!\odot\!\mathbf{G}_{\mathrm{err}})\grave{\mathbf{P}}_\mathrm{c}\mathbf{u}_{l}+\mathbf{n}_{\mathrm{r},l}, & \varsigma=\mathrm{SE\mbox{-}HD},\\[2mm]
\begin{aligned}[b]
&\bigl(\mathbf{D}_\mathrm{rc}^{T}\!\odot\!(\mathbf{G}_{\mathrm{r}}-\widehat{\mathbf{G}}_{\mathrm{r}})\bigr)
\mathbf{W}_{\mathrm{c},l}\mathbf{P}_{\mathrm{c}}\mathbf{s}_{l}\\
&+\beta_{\mathrm{R}}\sqrt{\tfrac{P_{\mathrm{r}}}{M_{\mathrm{r}}}}
\bigl(\mathbf{H}_\mathrm{R2I}-\widehat{\mathbf{H}}_\mathrm{R2I}\bigr)\mathbf{x}_{l}\\
&+\bigl(\grave{\mathbf{D}}_{\mathrm{ru}}\!\odot\!
(\mathbf{G}_{\mathrm{r}}^{(\mathrm{UL})}-\widehat{\mathbf{G}}_{\mathrm{r}}^{(\mathrm{UL})})\bigr)
\grave{\mathbf{P}}_{\mathrm c}\hat{\mathbf{u}}_{l}
+\mathbf{n}_{\mathrm{r},l},
\end{aligned} & \varsigma=\mathrm{SE\mbox{-}FD}.
\end{cases}
\label{eq:SE_radar_residual}
\vspace{-0.02 in}
\end{equation}
with $(\grave{\mathbf{D}}_{\mathrm{ru}}\!\odot\!\mathbf{G}_{\mathrm{err}})\grave{\mathbf{P}}_\mathrm{c}\mathbf{u}_{l}$ denoting the HD interference residual after IC of the UL-induced interference and $\mathbf{G}_{\mathrm{err}}\triangleq \mathbf{G}_{\mathrm{r}}^{(\mathrm{UL})}-\widehat{\mathbf{G}}_{\mathrm{r}}^{(\mathrm{UL})}\in \mathbb{C}^{M_\mathrm{r}\times K_{\mathrm{U}}}$ the channel-estimation-error matrix. Modelling the HD errors as i.i.d. $\mathcal{CN}(0,2\sigma_{\mathrm{RH},\mathrm{err}_1}^{2})$ and the three FD error matrices as i.i.d. $\mathcal{CN}(0,2\sigma_{\mathrm{RF},\mathrm{err}_1}^{2})$, the CLT gives $\boldsymbol{\omega}_{\mathrm{r}}^{(\varsigma)}\sim\mathcal{CN}\bigl(\mathbf{0},\,2\sigma_{\omega,\mathrm{r}}^{2(\varsigma)}\mathbf{I}_{M_{\mathrm{r}}}\bigr)$, with 
\vspace{-0.05 in}
\begin{equation}
\sigma_{\omega,\mathrm{r}}^{2(\varsigma)}=
\begin{cases}
\sigma_{\mathrm{RH},\mathrm{err}_1}^{2}\tfrac{1}{M_{\mathrm r}K_{\mathrm{U}}}\|\grave{\mathbf{D}}_{\mathrm{ru}}\|_F^2\operatorname{tr}(\grave{\mathbf{P}}_{\mathrm c}\grave{\mathbf{P}}_{\mathrm c}^H)+\sigma_n^2, & \varsigma=\mathrm{SE\mbox{-}HD},\\[2mm]
\begin{aligned}[b]
&\sigma_{\mathrm{RF},\mathrm{err}_1}^{2}\Bigl(\dfrac{1}{M_{\mathrm r}}\|\mathbf{D}_\mathrm{rc}\|_{F}^{2}\sigma_{\mathrm{w}}^2P_\mathrm{c}\\
&+\beta_{\mathrm{R}}^2\tfrac{P_{\mathrm{r}}}{M_{\mathrm{r}}}\operatorname{tr}(\mathbf{R}_{x})\\
&+\tfrac1{M_{\mathrm r}K_{\mathrm{U}}}\|\grave{\mathbf{D}}_{\mathrm{ru}}\|_{F}^{2}\operatorname{tr}(\grave{\mathbf{P}}_{\mathrm c}\grave{\mathbf{P}}_{\mathrm c}^H)\Bigr)+\sigma_n^2,
\end{aligned} & \varsigma=\mathrm{SE\mbox{-}FD}.
\end{cases}
\label{eq:SE_radar_var}
\vspace{-0.02 in}\end{equation}
Thus, in FD the residual power grows with the energies of the DL precoder, the radar waveform, and the detected UL symbols, each scaled by the common estimation-error variance. \label{r39b}

\vspace{-0.05 in}
\subsection{Shared Deployment}
In SH deployment, all $M$ APs are jointly utilised for sensing and communication: the CPU forms $T$ radar beams while simultaneously serving $K$ single-antenna UEs in the DL and UL. The transmit signal generated at the CPU is therefore a superposition of the radar and communication waveforms,
\vspace{-0.05 in}\begin{equation}
\dot{\mathbf{W}}_{\mathrm{c},l}\:\mathbf{P}_\mathrm{c}\:\mathbf{s}_{l}+\sqrt{\tfrac{P_{\mathrm{r}}}{M}}\,\dot{\mathbf{x}}_{l}, \label{eq:shh}
\vspace{-0.02 in}\end{equation}
where $\dot{\mathbf{x}}_{l}=\dot{\mathbf{W}}_{\mathrm{r},l}\,\dot{\mathbf{P}}_\mathrm{r}\,\boldsymbol{\phi}_{l}$ is the SH deployment radar waveform, with $\dot{\mathbf{W}}_{\mathrm{r},l}\in \mathbb{C}^{M\times T}$ following the structure of \eqref{eq:2}, $\dot{\mathbf{W}}_{\mathrm{c},l}\in \mathbb{C}^{M\times K_{\mathrm{D}}}$ the ZF precoder constructed as in the SE case from the composite channel $\dot{\mathbf{H}}+\dot{\mathbf{F}}$, and $\dot{\mathbf{P}}_\mathrm{r}=\mathrm{diag}(\sqrt{P_{\mathrm{r},1}/P_\mathrm{r}},\cdots,\sqrt{P_{\mathrm{r},T}/P_\mathrm{r}})$; the corresponding SH waveform covariance is $\dot{\mathbf{R}}_{x}\triangleq\frac{1}{L}\sum_{l=1}^{L}\dot{\mathbf{x}}_{l}\dot{\mathbf{x}}_{l}^{H}\in\mathbb{C}^{M\times M}$. The distributed steering vector $\dot{\mathbf{a}}_{t}\in\mathbb{C}^{M\times 1}$ and array response $\dot{\mathbf{A}}_t = \dot{\mathbf{a}}_{t}\dot{\mathbf{a}}_{t}^{T}$ are defined analogously to \eqref{eq:dist_steering}, with AP-target distances $\dot{d}_{\mathrm{r},t,m}$, $m=1,\ldots,M$. Under HD, transmission and reception follow the TDD schedule over the full SH aperture; under FD, all APs simultaneously transmit the composite waveform and receive UL signals and echoes within snapshot $l$. Since the APs share receive hardware, we set $\beta_{\mathrm{AP}}=\beta_{\mathrm{R}}\triangleq\beta$.

\subsubsection{Communication System: Downlink}
Under HD, the received signal $\dot{\mathbf{y}}_{l}^{(\mathrm{DL,HD})} \in \mathbb{C}^{K_{\mathrm{D}} \times 1}$ at the DL UEs is
\vspace{-0.04 in}
\begin{align}
\dot{\mathbf{y}}_{l}^{(\mathrm{DL,HD})}
&=
(\dot{\mathbf{D}}_\mathrm{c}\!\odot\!(\dot{\mathbf{H}}+\dot{\mathbf{F}}))
\dot{\mathbf{W}}_{\mathrm{c},l}\mathbf{P}_\mathrm{c}\mathbf{s}_{l}
\nonumber\\
&\quad+
\sqrt{\frac{P_{\mathrm{r}}}{M}}
(\dot{\mathbf{D}}_\mathrm{c}\!\odot\!(\dot{\mathbf{H}}+\dot{\mathbf{F}}))
\dot{\mathbf{x}}_{l}
+
\mathbf{n}_{l},
\label{eq:1sh}
\vspace{-0.02 in}
\end{align}
where the first term is the desired DL signal and the second is the radar-induced interference and clutter. Here, $\dot{\mathbf{H}}\in \mathbb{C}^{K_{\mathrm{D}}\times M}$ is the AP--UE channel and $\dot{\mathbf{F}}\in \mathbb{C}^{K_{\mathrm{D}}\times M}$ the clutter from APs to DL UEs, both are flat Rayleigh fading, with composite (direct-plus-clutter) channel $\dot{\mathbf{H}}+\dot{\mathbf{F}}$; the noise is $\mathbf{n}_l\sim\mathcal{CN}(\mathbf{0},2\sigma_n^2\mathbf{I}_{K_{\mathrm{D}}})$ and the large-scale fading is $\dot{\mathbf{D}}_{\mathrm{c}}\in \mathbb{R}^{K_{\mathrm{D}}\times M}$, $[\dot{\mathbf{D}}_{\mathrm{c}}]_{k,m}=\dot{d}_{\mathrm{c},k,m}^{-\eta/2}$. Under FD, the DL UEs receive while the UL UEs transmit, so the received signal $\dot{\mathbf{y}}_{l}^{(\mathrm{DL,FD})}$ contains extra UE-to-UE interference,
\vspace{-0.04 in}
\begin{align}
\!\!\dot{\mathbf{y}}_{l}^{(\mathrm{DL,FD})}
&=
(\dot{\mathbf{D}}_\mathrm{c}\!\odot\!(\dot{\mathbf{H}}+\dot{\mathbf{F}}))
\dot{\mathbf W}_{\mathrm c,l}\mathbf P_{\!\mathrm{c}}\mathbf s_l
\nonumber\\
&+
\sqrt{\tfrac{P_{\mathrm{r}}}{M}}
(\dot{\mathbf{D}}_\mathrm{c}\!\odot\!(\dot{\mathbf{H}}+\dot{\mathbf{F}}))
\dot{\mathbf x}_l
+
(\mathbf{D}_{\mathrm{u}}\!\odot\!\mathbf{H}_{\mathrm{u}})
\grave{\mathbf{P}}_{\mathrm{c}}\mathbf{u}_{l}
+
\mathbf n_l ,
\label{eq:SHFD_DL_raw}
\vspace{-0.02 in}
\end{align}
with $\dot{\mathbf{D}}_{\mathrm{c}}$ as above and the UE-to-UE pathloss $\mathbf{D}_{\mathrm{u}}$ and channel $\mathbf{H}_{\mathrm{u}}$ as defined for the SE case in \eqref{eq:DL_SEFD}. As in the SE case, the aggregate interference-plus-noise at the $k$-th DL UE comprises the clutter and radar interference, the UE-to-UE interference in the FD paradigm only, and the noise, taken from \eqref{eq:1sh} under HD and from \eqref{eq:SHFD_DL_raw} under FD. By the CLT it forms the zero-mean, circularly-symmetric complex Gaussian variable $\omega_{\mathrm{d},k}^{(\varsigma)}\sim\mathcal{CN}\bigl(0,2\sigma_{\omega,\mathrm{d},k}^{2(\varsigma)}\bigr)$, with per-dimension variance\label{r15a}\vspace{-0.04 in}

\vspace{-0.05 in}
\begin{equation}
\sigma_{\omega,\mathrm{d},k}^{2(\varsigma)}=
\begin{cases}
\frac{P_{\mathrm{r}}\operatorname{tr}(\dot{\mathbf{R}}_{x})}{M^{2}}\sum_{m=1}^{M}\dot{d}_{\mathrm{c},k,m}^{-\eta}(\sigma_{H}^{2}+\sigma_{F}^{2})+\sigma_{n}^{2}, & \varsigma=\mathrm{SH\mbox{-}HD},\\[2mm]
\begin{aligned}[b]
&\frac{P_{\mathrm{r}}\operatorname{tr}(\dot{\mathbf{R}}_{x})}{M^{2}}\sum_{m=1}^{M}\dot{d}_{\mathrm{c},k,m}^{-\eta}(\sigma_{H}^{2}+\sigma_{F}^{2})\\
&+\sum_{j=1}^{K_{\mathrm{U}}}\grave{P}_{\mathrm{c},j}d_{\mathrm{u},k,j}^{-\eta}\sigma_{\mathrm{u}}^{2}+\sigma_{n}^{2},
\end{aligned} & \varsigma=\mathrm{SH\mbox{-}FD}.
\end{cases}
\label{eq:SH_DL_var}
\vspace{-0.02 in}\end{equation}

\subsubsection{Communication System: Uplink}
During the UL phase, the CPU collects the signals received at all APs. Here, $\grave{\mathbf{H}}\in\mathbb{C}^{M \times K_{\mathrm{U}}}$ is the UL channel with i.i.d. entries $\mathcal{CN}(0,2\sigma_H^2)$, $\dot{\mathbf{H}}_{\mathrm{R2C}}\in\mathbb{C}^{M\times M}$ is the backscattering channel of the composite waveform with i.i.d. entries $\mathcal{CN}(0,2\sigma_H^2)$, and the large-scale fading matrices $\grave{\mathbf{D}}_{\mathrm{c}}\in \mathbb{R}^{M\times K_{\mathrm{U}}}$ and $\dot{\mathbf{D}}_{\mathrm{rc}}\in \mathbb{R}^{M\times M}$ capture the UL-UE-to-AP and AP-to-AP backscatter links, respectively, with $[\grave{\mathbf{D}}_{\mathrm{c}}]_{m,k}=\grave{d}_{\mathrm{c},m,k}^{-\eta/2}$ and $[\dot{\mathbf{D}}_{\mathrm{rc}}]_{m,n}=\dot{d}_{\mathrm{rc},m,n}^{-\eta/2}$. Under HD, the received signal is
\vspace{-0.05 in}
\begin{equation}
\dot{\mathbf{y}}_l^{(\mathrm{UL,HD})}
=
(\grave{\mathbf{D}}_\mathrm{c}\!\odot\!\grave{\mathbf{H}})
\grave{\mathbf{P}}_\mathrm{c}\mathbf{u}_{l}
+
(\dot{\mathbf{D}}_\mathrm{rc}\!\odot\!\dot{\mathbf{H}}_{\mathrm{R2C}})
(\dot{\mathbf{W}}_{\mathrm{c},l}\mathbf{P}_\mathrm{c}\mathbf{s}_{l}
+\sqrt{\tfrac{P_{\mathrm{r}}}{M}}\dot{\mathbf{x}}_{l})
+
\mathbf{n}_l,
\vspace{-0.02 in}
\end{equation}
whereas under FD it is additionally impaired by SI from the composite transmitted waveform,
\vspace{-0.04 in}
\begin{align}
\dot{\mathbf{y}}_l^{(\mathrm{UL,FD})}
&=
(\grave{\mathbf{D}}_\mathrm{c}\!\odot\!\grave{\mathbf{H}})
\grave{\mathbf{P}}_\mathrm{c}\mathbf{u}_{l}
+
(\dot{\mathbf{D}}_\mathrm{rc}\!\odot\!\dot{\mathbf{H}}_{\mathrm{R2C}})
(\dot{\mathbf{W}}_{\mathrm{c},l}\mathbf{P}_\mathrm{c}\mathbf{s}_{l}
+\sqrt{\tfrac{P_{\mathrm{r}}}{M}}\dot{\mathbf{x}}_{l})
\nonumber\\
&\quad+
\beta_{\mathrm{AP}}\dot{\mathbf{H}}_{\mathrm{SI}}
\left(\dot{\mathbf{W}}_{\mathrm{c},l}\mathbf{P}_\mathrm{c}\mathbf{s}_{l}
+\sqrt{\tfrac{P_{\mathrm{r}}}{M}}\dot{\mathbf x}_l\right)
+
\mathbf{n}_l,
\label{eq:SHFD_UL_raw}
\vspace{-0.02 in}
\end{align}
where $\dot{\mathbf{H}}_\mathrm{SI}\in\mathbb{C}^{M\times M}$ is the effective SI channel with i.i.d. entries $\mathcal{CN}(0,2\sigma_H^2)$. Since the composite waveform is generated at the CPU, the backscattered-signal and SI terms are digitally suppressed via MMSE estimates; after cancellation, the UL received signal for both duplexities takes the unified form
      \vspace{-0.05 in}
\begin{equation}
\mathbf{y}_l^{(\mathrm{UL},\varsigma)}
=
(\grave{\mathbf{D}}_\mathrm{c}\!\odot\!\grave{\mathbf{H}})
\grave{\mathbf{P}}_\mathrm{c}\mathbf{u}_{l}
+
\boldsymbol{\omega}_\mathrm{c}^{(\varsigma)}
+
\mathbf{n}_l,
\label{eq:shared_UL_afterIC}
\vspace{-0.02 in}
\end{equation}
where the post-IC residual interference is in \eqref{eq:SH_UL_residual}.

\begin{table*}
\begin{minipage}{1\textwidth}
\begin{align}
\boldsymbol{\omega}_{\mathrm{c}}^{(\varsigma)}=
\begin{cases}
\bigl(\dot{\mathbf{D}}_\mathrm{rc}\!\odot\!(\dot{\mathbf{H}}_\mathrm{R2C}-\widehat{\dot{\mathbf{H}}}_\mathrm{R2C})\bigr)
(\dot{\mathbf{W}}_{\mathrm{c},l}\mathbf{P}_\mathrm{c}\mathbf{s}_{l}+\sqrt{\tfrac{P_{\mathrm{r}}}{M}}\dot{\mathbf{x}}_{l}),
 & \varsigma=\mathrm{SH\mbox{-}HD},\\
\bigl(\dot{\mathbf{D}}_\mathrm{rc}\!\odot\!(\dot{\mathbf{H}}_\mathrm{R2C}-\widehat{\dot{\mathbf{H}}}_\mathrm{R2C})\bigr)(\dot{\mathbf{W}}_{\mathrm{c},l}\mathbf{P}_\mathrm{c}\mathbf{s}_{l}+\sqrt{\tfrac{P_{\mathrm{r}}}{M}}\dot{\mathbf{x}}_{l})
+\beta_{\mathrm{AP}}\bigl(\dot{\mathbf{H}}_{\mathrm{SI}}-\widehat{\dot{\mathbf{H}}}_{\mathrm{SI}}\bigr)
\times(\dot{\mathbf{W}}_{\mathrm{c},l}\mathbf{P}_\mathrm{c}\mathbf{s}_{l}+\sqrt{\tfrac{P_{\mathrm{r}}}{M}}\dot{\mathbf{x}}_{l}),& \varsigma=\mathrm{SH\mbox{-}FD}.
\end{cases}
\label{eq:SH_UL_residual}
\end{align}
\vspace{-0.15in}
\medskip
\hrule
\vspace{-0.24in}
\end{minipage}
\end{table*}

Modelling the estimation errors in the HD and FD cases as i.i.d. $\mathcal{CN}(0,2\sigma_{\mathrm{UH},\mathrm{err}_2}^{2})$ and $\mathcal{CN}(0,2\sigma_{\mathrm{UF},\mathrm{err}_2}^{2})$, respectively, each element of $\boldsymbol{\omega}_{\mathrm{c}}^{(\varsigma)}$ combines $M$ independent zero-mean terms, so by the CLT $\boldsymbol{\omega}_{\mathrm{c}}^{(\varsigma)}\sim\mathcal{CN}\!\bigl(\mathbf{0},2\sigma_{\omega,\mathrm{c}}^{2(\varsigma)}\mathbf{I}_{M}\bigr)$, with
\vspace{-0.05 in}
\begin{equation}
\sigma_{\omega,\mathrm{c}}^{2(\varsigma)}=
\begin{cases}
\begin{aligned}[b]
&\dfrac{\sigma_{\mathrm{UH},\mathrm{err}_2}^{2}}{M^2}\|\dot{\mathbf{D}}_\mathrm{rc}\|_{F}^{2}\\
&\times\bigl(\sigma_{\dot{\mathrm{w}}}^2 M P_\mathrm{c} + \tfrac{P_{\mathrm{r}}}{M}\operatorname{tr}(\dot{\mathbf{R}}_{x})\bigr),
\end{aligned} & \varsigma=\mathrm{SH\mbox{-}HD},\\[2mm]
\begin{aligned}[b]
&\sigma_{\mathrm{UF},\mathrm{err}_2}^{2}\bigl(\dfrac{\|\dot{\mathbf D}_{\mathrm{rc}}\|_{F}^{2}}{M^2}+\beta_{\mathrm{AP}}^{2}\bigr)\\
&\times\bigl(\sigma_{\dot{\mathrm{w}}}^{2}MP_{\mathrm c}+\tfrac{P_{\mathrm r}}{M}\operatorname{tr}(\dot{\mathbf R}_x)\bigr),
\end{aligned} & \varsigma=\mathrm{SH\mbox{-}FD}.
\end{cases}
\label{eq:SH_UL_var}
\vspace{-0.02 in}\end{equation}
The CPU then applies the ZF combiner $\mathbf{G}_{\mathrm{c}}$ defined for the SE case, now formed from the SH UL channel $\grave{\mathbf{H}}\in\mathbb{C}^{M\times K_{\mathrm{U}}}$, to $\mathbf{y}_{l}^{(\mathrm{UL},\varsigma)}$ to detect the UL symbols.

\subsubsection{Radar System}
Under HD, the radar return signal for beam $t$ under $\mathcal{H}_q$ is
\vspace{-0.04 in}
\begin{align}
\dot{\mathbf{y}}_{\mathrm{r},t,l|\mathcal{H}_{q}}^{(\mathrm{HD})}
&=
\alpha_t
(\dot{\mathbf{D}}_{\mathrm{r},t}\!\odot\!\dot{\mathbf{A}}_t)
\left(
\sqrt{\frac{P_{\mathrm{r},t}}{M}}
\dot{\mathbf{w}}_{\mathrm{r},t,l}
+
\dot{\mathbf{W}}_{\mathrm{c},l}\mathbf{P}_\mathrm{c}\mathbf{s}_{l}
\right)q
\nonumber\\
&\quad+
(\grave{\mathbf{D}}_\mathrm{c}\!\odot\!\dot{\mathbf{G}}_\mathrm{r}^{(\mathrm{UL})})
\grave{\mathbf{P}}_\mathrm{c}\mathbf{u}_{l}
+
\dot{\mathbf{n}}_{\mathrm{r},l},
\label{eq:shrad}
\vspace{-0.02 in}
\end{align}
where $\dot{\mathbf{G}}_\mathrm{r}^{(\mathrm{UL})}\in\mathbb{C}^{M\times K_{\mathrm{U}}}$ is the channel from the $K_{\mathrm{U}}$ UEs to the $M$ APs, the pathloss vector is $\dot{\mathbf{d}}_{\mathrm{r},t}\triangleq\bigl[\dot{d}_{\mathrm{r},t,1}^{-\eta/2},\ldots,\dot{d}_{\mathrm{r},t,M}^{-\eta/2}\bigr]^T$ with rank-one matrix $\dot{\mathbf{D}}_{\mathrm{r},t}\triangleq \dot{\mathbf{d}}_{\mathrm{r},t}\dot{\mathbf{d}}_{\mathrm{r},t}^{T}\in\mathbb{R}^{M\times M}$, and the first parenthesised term is the transmitted radar-plus-communication signal reflected back by the target, while the second is the UL interference. Under FD, the radar APs receive echoes while transmitting the composite waveform and receiving UL transmissions, so the radar return signal additionally contains AP-to-AP backscattered signal and clutter and composite-waveform SI,
\vspace{-0.04 in}
\begin{align}
\dot{\mathbf{y}}_{\mathrm{r},t,l|\mathcal{H}_{q}}^{(\mathrm{FD})}
&=
\alpha_t
(\dot{\mathbf{D}}_{\mathrm{r},t}\!\odot\!\dot{\mathbf{A}}_t)
\left(
\sqrt{\frac{P_{\mathrm{r},t}}{M}}
\dot{\mathbf{w}}_{\mathrm{r},t,l}
+
\dot{\mathbf{W}}_{\mathrm{c},l}\mathbf{P}_\mathrm{c}\mathbf{s}_{l}
\right)q
\nonumber\\&\!\!\!\!\!\!\!+
(\dot{\mathbf{D}}_{\mathrm{rc}}\!\odot\!\dot{\mathbf{G}}_\mathrm{r})
\left(
\sqrt{\frac{P_{\mathrm{r}}}{M}}\dot{\mathbf{x}}_{l}
+
\dot{\mathbf{W}}_{\mathrm{c},l}\mathbf{P}_\mathrm{c}\mathbf{s}_{l}
\right)\!\!+\!
(\grave{\mathbf D}_\mathrm{c}\!\odot\!
\dot{\mathbf{G}}_\mathrm{r}^{(\mathrm{UL})})
\grave{\mathbf P}_{\mathrm c}\mathbf u_l
\nonumber\\
&\!\!\!\!\!\!\!+
\beta_{\mathrm{R}}\dot{\mathbf{H}}_\mathrm{SI}\!
\left(
\sqrt{\tfrac{P_{\mathrm r}}{M}}\dot{\mathbf x}_l
+
\dot{\mathbf W}_{\mathrm c,l}\mathbf P_\mathrm{c}\mathbf s_l
\right)
+
\dot{\mathbf n}_{\mathrm r,l},
\label{eq:shrad1}
\vspace{-0.02 in}
\end{align}
where $\dot{\mathbf{G}}_\mathrm{r} \in \mathbb{C}^{M \times M}$ is the AP-to-AP backscatter (clutter) channel with i.i.d. entries $\mathcal{CN}(0,2\sigma_H^2)$, and the parenthesised terms are i) the backscattered return, ii) the clutter towards the APs, iii) the UL interference, and iv) the composite-waveform SI. Since the composite waveform is known at the CPU and $\hat{\mathbf{u}}_l$ is recovered after UL decoding, the terms cancelled by IC are the UL-induced interference in the HD case, together with the clutter and SI terms in the FD case, so that the post-IC radar return signal for both duplexities takes the unified form
\vspace{-0.05 in}
\begin{equation}
\mathbf{y}_{\mathrm{r},t,l|\mathcal{H}_{q}}^{(\varsigma)} =\alpha_t\sqrt{\frac{
P_{\mathrm{r},t}}{M}}(\dot{\mathbf{D}}_{\mathrm{r},t}\!\odot\!\dot{\mathbf{A}}_t)
\dot{\mathbf{w}}_{\mathrm{r},t,l}q+\boldsymbol{\omega}_\mathrm{r}^{(\varsigma)},
\label{eq:SH_radar_afterIC}
\vspace{-0.02 in}\end{equation}
where, following the SE deployment convention of \eqref{eq:SE_radar_residual}, the residual interference-plus-noise is shown in \eqref{eq:SH_radar_residual}, where $\mathbf{G}_{\mathrm{err_1}}\triangleq\dot{\mathbf{G}}_\mathrm{r}^{(\mathrm{UL})}-\widehat{\dot{\mathbf{G}}}_\mathrm{r}^{(\mathrm{UL})}\in\mathbb{C}^{M\times K_{\mathrm{U}}}$ and $\mathbf{G}_{\mathrm{err_2}}\in\mathbb{C}^{M\times M}$ are the channel-estimation-error matrices of the UL-induced interference and of the target-reflected communication waveform, respectively, with i.i.d. entries $\mathcal{CN}(0,2\sigma_{\mathrm{RH},\mathrm{err}_2}^{2})$ under HD and $\mathcal{CN}(0,2\sigma_{\mathrm{RF},\mathrm{err}_2}^{2})$ under FD. Under FD the residual instead comprises the imperfectly cancelled clutter/backscattering ($\dot{\mathbf G}_\mathrm{r}-\widehat{\dot{\mathbf G}}_\mathrm{r}$), SI ($\dot{\mathbf H}_\mathrm{SI}-\widehat{\dot{\mathbf H}}_\mathrm{SI}$), and residual UL interference ($\dot{\mathbf G}_\mathrm{r}^{(\mathrm{UL})}-\widehat{\dot{\mathbf G}}_\mathrm{r}^{(\mathrm{UL})}$), each multiplying the composite transmitted signal and modelled as i.i.d. $\mathcal{CN}(0,2\sigma_{\mathrm{RF},\mathrm{err}_2}^{2})$. By the CLT, $\boldsymbol{\omega}_\mathrm{r}^{(\varsigma)}\sim\mathcal{CN}\bigl(\mathbf 0,2\sigma_{\omega,\mathrm{r}}^{2(\varsigma)}\mathbf I_M\bigr)$, where the variance $\sigma_{\omega,\mathrm{r}}^{2(\varsigma)}$ is shown in \eqref{eq:SH_radar_var}.
\begin{table*}
\begin{minipage}{1\textwidth}
\begin{align}
&\boldsymbol{\omega}_{\mathrm{r}}^{(\varsigma)}=
\begin{cases}
(\dot{\mathbf{D}}_{\mathrm{r},t}\!\odot\!\mathbf{G}_\mathrm{err_2})
\dot{\mathbf{W}}_{\mathrm{c},l}\mathbf{P}_\mathrm{c}\mathbf{s}_{l}
+
(\grave{\mathbf{D}}_\mathrm{c}\!\odot\!\mathbf{G}_\mathrm{err_1})
\grave{\mathbf{P}}_\mathrm{c}\mathbf{u}_{l}
+
\dot{\mathbf{n}}_{\mathrm{r},l},
& \hspace{1.52 in} \varsigma=\mathrm{SH\mbox{-}HD},\\[2mm]
\begin{aligned}[b]
&(\dot{\mathbf{D}}_{\mathrm{r},t}\!\odot\!\mathbf{G}_\mathrm{err_2})
\dot{\mathbf{W}}_{\mathrm{c},l}\mathbf{P}_\mathrm{c}\mathbf{s}_{l}
+
\bigl(\dot{\mathbf{D}}_{\mathrm{rc}}\!\odot\!
(\dot{\mathbf{G}}_\mathrm{r}\!-\!\widehat{\dot{\mathbf{G}}}_\mathrm{r})\bigr)
\bigl(\sqrt{\tfrac{P_{\mathrm{r}}}{M}}\dot{\mathbf{x}}_{l}
+\dot{\mathbf{W}}_{\mathrm{c},l}\mathbf{P}_\mathrm{c}\mathbf{s}_{l}\bigr)\\
&+\beta_{\mathrm{R}}
\bigl(\dot{\mathbf{H}}_\mathrm{SI}\!-\!\widehat{\dot{\mathbf{H}}}_\mathrm{SI}\bigr)
\bigl(\sqrt{\tfrac{P_{\mathrm{r}}}{M}}\dot{\mathbf{x}}_{l}
+\dot{\mathbf{W}}_{\mathrm{c},l}\mathbf{P}_\mathrm{c}\mathbf{s}_{l}\bigr)
+
\bigl(\grave{\mathbf{D}}_\mathrm{c}\!\odot\!
(\dot{\mathbf{G}}_\mathrm{r}^{(\mathrm{UL})}
\!-\!\widehat{\dot{\mathbf{G}}}_\mathrm{r}^{(\mathrm{UL})})\bigr)
\grave{\mathbf{P}}_\mathrm{c}\hat{\mathbf{u}}_{l}
+\dot{\mathbf{n}}_{\mathrm{r},l},
\end{aligned}
& \hspace{1.52 in} \varsigma=\mathrm{SH\mbox{-}FD}.
\end{cases}
\label{eq:SH_radar_residual}
\\ \nonumber \\ &\sigma_{\omega,\mathrm{r}}^{2(\varsigma)}=
\begin{cases}
\sigma_{\mathrm{n}}^2+\sigma_{\mathrm{RH},\mathrm{err}_2}^{2}\bigl(\tfrac{1}{M}\|\dot{\mathbf{D}}_{\mathrm{r},t}\|_F^2\sigma_{\mathrm{\dot{w}}}^2P_\mathrm{c}+\tfrac1{K_{\mathrm{U}}M}\|\grave{\mathbf{D}}_\mathrm{c}\|_F^2\operatorname{tr}(\grave{\mathbf{P}}_\mathrm{c}\grave{\mathbf{P}}_\mathrm{c}^H)\bigr), &\hspace{0.1 in} \varsigma=\mathrm{SH\mbox{-}HD},\\[3mm]
\begin{aligned}[b]
&\sigma_{n}^{2}+\sigma_{\mathrm{RF},\mathrm{err}_2}^{2}\Bigl[\tfrac{1}{M}\|\dot{\mathbf{D}}_{\mathrm{r},t}\|_F^2\sigma_{\mathrm{\dot{w}}}^2P_\mathrm{c}+\bigl(\tfrac{1}{M^2}\|\dot{\mathbf{D}}_{\mathrm{rc}}\|_F^2+\beta_{\mathrm{R}}^{2}\bigr)\bigl(\sigma_{\dot{\mathrm{w}}}^{2}M P_{\mathrm c}+\tfrac{P_{\mathrm r}}{M}\operatorname{tr}(\dot{\mathbf R}_x)\bigr)+\tfrac1{K_{\mathrm{U}}M}\|\grave{\mathbf D}_{\mathrm c}\|_{F}^{2}\operatorname{tr}(\grave{\mathbf{P}}_\mathrm{c}\grave{\mathbf{P}}_\mathrm{c}^H)\Bigr].
\end{aligned} & \hspace{0.1 in} \varsigma=\mathrm{SH\mbox{-}FD}.
\end{cases}
\label{eq:SH_radar_var}
\end{align}
\vspace{-0.15in}
\medskip
\hrule
\vspace{-0.24in}
\end{minipage}
\end{table*}
\label{r39c}
\section{Radar Detection and Estimation Framework}

This section develops the GLRT-based detection framework and the associated detection-probability metrics for the radar subsystem across all operating configurations, using the configuration index $\varsigma\in\mathcal{S}$ and the superscript-$(\varsigma)$ convention of Section~II. This formulation is required because the radar KLD introduced in Section~IV is not an isolated information-theoretic quantity but is tied to the radar hypothesis test: separating the target-present and target-absent hypotheses calls for a decision rule, for which the well-established GLRT is invoked to form the radar test statistic and its distributions under each hypothesis. These distributions yield the false-alarm and detection probabilities, and allow the radar KLD to be interpreted operationally through the miss-detection exponent of Stein's lemma, thereby placing sensing on the same distinguishability scale as communication and underpinning the unified comparison that is the objective of this work.

\subsection{Unified GLRT Formulation}
After collecting $L$ snapshots, the radar return signal matrix for the $t$-th spatial--range--Doppler bin can be expressed uniformly across all configurations as,
\vspace{-0.05 in}\begin{equation}
\mathbf{Y}_{\mathrm{r},t|\mathcal{H}_{q}} =
\sqrt{\frac{P_{\mathrm{r},t}}{M^{(\varsigma)}}}\,
\alpha_t\,
(\mathbf{D}_{\mathrm{r},t}^{(\varsigma)}\!\odot\!\mathbf{A}_t^{(\varsigma)})\mathbf{W}_{\mathrm{r},t}^{(\varsigma)}\,q
+ \boldsymbol{\Omega}^{(\varsigma)},
\label{eq:signal_matrix}
\vspace{-0.02 in}\end{equation}
where $\mathbf{W}_{\mathrm{r},t}^{(\varsigma)} \in \mathbb{C}^{M^{(\varsigma)} \times L}$ collects the $L$ radar precoding vectors $\mathbf{w}_{\mathrm{r},t,l}^{(\varsigma)}$ of the $t$-th beam, and $M^{(\varsigma)} \in \{M_{\mathrm{r}}, M\}$ is the number of APs used for radar, namely $M_{\mathrm{r}}$ in SE and all $M$ in SH deployment. The rank-one pathloss matrix $\mathbf{D}_{\mathrm{r},t}^{(\varsigma)}$ and array response $\mathbf{A}_t^{(\varsigma)}$ of bin $t$, both $M^{(\varsigma)}\!\times\! M^{(\varsigma)}$, take the SE forms $(\mathbf{D}_{\mathrm{r},t},\mathbf{A}_t)$ or SH forms $(\dot{\mathbf{D}}_{\mathrm{r},t},\dot{\mathbf{A}}_t)$ of Section~II and are fixed by the known AP locations and bin geometry, leaving the complex reflection coefficient $\alpha_t$ as the only unknown, estimated by ML. With the rank-one factorisations $\mathbf{D}_{\mathrm{r},t}^{(\varsigma)}=\mathbf{d}_{\mathrm{r},t}^{(\varsigma)}(\mathbf{d}_{\mathrm{r},t}^{(\varsigma)})^{T}$ and $\mathbf{A}_t^{(\varsigma)}=\mathbf{a}_t^{(\varsigma)}(\mathbf{a}_{t}^{(\varsigma)})^{T}$, where $\mathbf{d}_{\mathrm{r},t}^{(\varsigma)}\in\mathbb{R}^{M^{(\varsigma)}\times 1}$ and $\mathbf{a}_t^{(\varsigma)}\in\mathbb{C}^{M^{(\varsigma)}\times 1}$, the Hadamard product collapses to the rank-one form $(\mathbf{d}_{\mathrm{r},t}^{(\varsigma)}\odot\mathbf{a}_t^{(\varsigma)})(\mathbf{d}_{\mathrm{r},t}^{(\varsigma)}\odot\mathbf{a}_t^{(\varsigma)})^{T}$. Since the bin geometry is fixed and known, the CPU phase-aligns the receive branches through the unitary rotation $\boldsymbol{\Theta}_t\triangleq\mathrm{diag}\bigl(\mathbf{a}_t^{(\varsigma)}\odot\mathbf{a}_t^{(\varsigma)}\bigr)^{H}$, which compensates the known round-trip phase of each branch and leaves the circularly symmetric residual statistically unchanged; absorbing $\boldsymbol{\Theta}_t$ into $\mathbf{Y}_{\mathrm{r},t|\mathcal{H}_{q}}$ in \eqref{eq:signal_matrix} without loss of generality, the target response takes the Hermitian rank-one form $\boldsymbol{\Theta}_t\bigl(\mathbf{D}_{\mathrm{r},t}^{(\varsigma)}\!\odot\!\mathbf{A}_t^{(\varsigma)}\bigr)=\mathbf{g}_{t}^{(\varsigma)}(\mathbf{g}_{t}^{(\varsigma)})^{H}$ with effective steering vector $\mathbf{g}_t^{(\varsigma)}\triangleq\mathbf{d}_{\mathrm{r},t}^{(\varsigma)}\odot\boldsymbol{\Theta}_t\mathbf{a}_t^{(\varsigma)}$, and $\mathbf{D}_{\mathrm{r},t}^{(\varsigma)}\!\odot\!\mathbf{A}_t^{(\varsigma)}$ denotes this phase-aligned response in the sequel. The residual matrix $\boldsymbol{\Omega}^{(\varsigma)}$ stacks the post-IC interference-plus-noise vectors, each column modelled as $\boldsymbol{\omega}^{(\varsigma)}_{l}\sim \mathcal{CN}\bigl(\mathbf 0,\,2\sigma_{\omega,\mathrm{r}}^{2(\varsigma)}\mathbf I_{M^{(\varsigma)}}\bigr)$ with configuration-dependent variance $\sigma_{\omega,\mathrm{r}}^{2(\varsigma)}$ from the case expressions of Section~II. Denoting the $l$-th column of $\mathbf{Y}_{\mathrm{r},t|\mathcal{H}_q}$ by $\mathbf{y}_{\mathrm{r},t,l|\mathcal{H}_{q}}^{(\varsigma)} \in \mathbb{C}^{M^{(\varsigma)}}$, the log-likelihood conditioned on $\alpha_t$ is,
\begin{multline}
\!\ln f\!\left(\mathbf{Y}_{\mathrm{r},t|\mathcal{H}_{1}};\alpha_t\right)
= -M^{(\varsigma)}L\ln(2\pi\sigma_{\omega,\mathrm{r}}^{2(\varsigma)}) \\
\!\!\!-\! \frac{1}{2\sigma_{\!\omega,\mathrm{r}}^{2(\varsigma)}}
\!\sum_{l=1}^{L}\!\left\|\mathbf{y}_{\mathrm{r},t,l|\mathcal{H}_{1}}^{(\varsigma)} \!\!-\!
\sqrt{\frac{P_{\mathrm{r},t}}{M^{(\varsigma)}}}
\alpha_t
(\mathbf{D}_{\mathrm{r},t}^{(\varsigma)}\!\odot\!\mathbf{A}_t^{(\varsigma)})\mathbf{w}_{\mathrm{r},t,l}^{(\varsigma)}\!\right\|^2\!\!\!.
\end{multline}
By expanding the squared norm and discarding terms that do not depend on
$\alpha_t$, a sufficient statistic for $\alpha_t$ is the target-present instance $\mathbf{E}_{t,1}$ of the per-hypothesis matrix,
\vspace{-0.05 in}\begin{equation}
\mathbf{E}_{t,q} = \frac{1}{L}\sum_{l=1}^{L}
\mathbf{y}_{\mathrm{r},t,l|\mathcal{H}_{q}}^{(\varsigma)}(\mathbf{w}_{\mathrm{r},t,l}^{(\varsigma)})^{H}.
\label{eq:sufficient_stat}
\vspace{-0.02 in}\end{equation}
Defining the transmit covariance $\mathbf{R}_t = \frac{1}{L}\sum_{l=1}^{L}
\mathbf{w}_{\mathrm{r},t,l}^{(\varsigma)}(\mathbf{w}_{\mathrm{r},t,l}^{(\varsigma)})^{H}$ and using the eigendecomposition $\mathbf{R}_t = \mathbf{U}\boldsymbol{\Lambda}\mathbf{U}^H$, vectorising the whitened statistic $\mathbf{E}_{t,1}\mathbf{U}\boldsymbol{\Lambda}^{-1/2}$ gives,
\vspace{-0.05 in}\begin{equation}
\mathbf{e}_{t,1} =
\sqrt{\frac{P_{\mathrm{r},t}}{M^{(\varsigma)}}}\,
\alpha_t\,\mathbf{d}_{\mathbf{w},t}
+ \tilde{\mathbf{n}},
\vspace{-0.02 in}\end{equation}
where $\mathbf{d}_{\mathbf{w},t} = \mathrm{vec}\!\left\{
(\mathbf{D}_{\mathrm{r},t}^{(\varsigma)}\!\odot\!\mathbf{A}_t^{(\varsigma)})\mathbf{U}\boldsymbol{\Lambda}^{1/2}\right\}$ is the equivalent array steering vector, expressed through the whitened waveform basis $\breve{\mathbf{w}}_{\mathrm{r},t,l} = \boldsymbol{\Lambda}^{-1/2}\mathbf{U}^H\mathbf{w}_{\mathrm{r},t,l}^{(\varsigma)}$, which satisfies $\frac{1}{L}\sum_{l=1}^{L}\breve{\mathbf{w}}_{\mathrm{r},t,l}\breve{\mathbf{w}}_{\mathrm{r},t,l}^{H}=\mathbf{I}$, and
$\tilde{\mathbf{n}} = \frac{1}{L}\mathrm{vec}\!\left\{
\sum_{l=1}^L \boldsymbol{\omega}^{(\varsigma)}_{l}
\,\breve{\mathbf{w}}_{\mathrm{r},t,l}^H\right\}
\sim \mathcal{CN}\!\left(\mathbf 0,\,\tfrac{2}{L}\sigma_{\omega,\mathrm{r}}^{2(\varsigma)}\mathbf I\right)$. The ML estimate of $\alpha_t$ is thus obtained from the least-squares problem,
\vspace{-0.05 in}\begin{equation}
\hat{\alpha}_t =
\arg\min_{\alpha_t}\,
\big\|\mathbf{e}_{t,1} - \sqrt{\tfrac{P_{\mathrm{r},t}}{M^{(\varsigma)}}}\,\alpha_t\mathbf{d}_{\mathbf{w},t}\big\|^2.
\label{eq:ml_estimates_vec}
\vspace{-0.02 in}\end{equation}
\subsection{Test Statistic and Asymptotic Form}

The GLRT is formulated as
\vspace{-0.05 in}\begin{equation}
\xi_t \overset{\mathcal{H}_1}{\underset{\mathcal{H}_0}{\gtrless}} \tau,
\qquad
\xi_t =
\ln\left(\frac{f(\mathbf{e}_{t,1};\hat{\alpha}_t,\mathcal{H}_1)}
{f(\mathbf{e}_{t,1};\mathcal{H}_0)}\right),
\label{eq:test_statistic_GLRT}
\vspace{-0.02 in}\end{equation}
where $\hat{\alpha}_t$ is the ML estimate obtained from \eqref{eq:ml_estimates_vec}. As $L\to\infty$, the ML estimator is asymptotically consistent ($\hat{\alpha}_t \overset{\text{asymp.}}{\longrightarrow} \alpha_t$). Substituting the asymptotic estimate into \eqref{eq:test_statistic_GLRT} and retaining the coherent matched-filter component reduces the test to the normalised energy of the combined statistic $\operatorname{tr}(\mathbf{E}_{t,q})=\frac{1}{L}\sum_{l=1}^{L}(\mathbf{w}_{\mathrm{r},t,l}^{(\varsigma)})^{H}\mathbf{y}_{\mathrm{r},t,l|\mathcal{H}_{q}}^{(\varsigma)}$,
\vspace{-0.05 in}\begin{equation}
\xi_t \;=\;
\frac{L}{\sigma_{\omega,\mathrm{r}}^{2(\varsigma)}\operatorname{tr}(\mathbf{R}_t)}
\left|
\sqrt{\frac{P_{\mathrm{r},t}}{M^{(\varsigma)}}}\,
\alpha_t\,
(\mathbf{g}_{t}^{(\varsigma)})^H\mathbf{R}_t\mathbf{g}_{t}^{(\varsigma)}
+ \breve{n}
\right|^2,
\label{eq:xi_asymptotic}
\vspace{-0.02 in}\end{equation}
where $\breve{n}\sim\mathcal{CN}\!\bigl(0,\tfrac{2}{L}\sigma_{\omega,\mathrm{r}}^{2(\varsigma)}\operatorname{tr}(\mathbf{R}_t)\bigr)$ is the scalar residual after coherent processing of the $L$ snapshots. Since the normalised squared magnitude of a complex Gaussian is Chi-squared distributed, the statistic in \eqref{eq:xi_asymptotic} follows $\xi_t \sim \chi_2^2(0)$ under $\mathcal{H}_0$ and $\xi_t \sim \chi_2^2(\lambda_t)$ under $\mathcal{H}_1$, where the noncentrality parameter is,
\vspace{-0.05 in}\begin{equation}
\lambda_t =
\frac{L|\alpha_t|^2P_{\mathrm{r},t}\,
\big|(\mathbf{g}_{t}^{(\varsigma)})^H\mathbf{R}_t\mathbf{g}_{t}^{(\varsigma)}\big|^2}
{\sigma_{\omega,\mathrm{r}}^{2(\varsigma)}\,M^{(\varsigma)}\,\operatorname{tr}(\mathbf{R}_t)}.
\label{eq:lambda_t}
\vspace{-0.02 in}\end{equation}
\subsection{False Alarm and Detection Probabilities}

Since $\xi_t\sim\chi_2^2(0)$ under $\mathcal{H}_0$, the false-alarm probability is $P_{\mathrm{FA}} =\Pr\{\xi_t > \tau \mid \mathcal{H}_0\}= \exp(-\tau/2)$. For a target level $P_{\mathrm{FA}}$, the threshold is
\vspace{-0.05 in}\begin{equation}
\tau = -2\ln(P_{\mathrm{FA}}).
\label{eq:threshold_setting}
\vspace{-0.02 in}\end{equation}
Under $\mathcal{H}_1$, the detection probability for the $t$-th bin is,
\vspace{-0.05 in}\begin{equation}
\!\!P_{\mathrm{D},t}
\!= \!\Pr\{\xi_t\! > \!\tau \!\mid\! \mathcal{H}_1\!\}\!=\! Q_{\chi_2^2}\bigl(\tau,\lambda_t\!\bigr)
\!= \!Q_1\bigl(\!\sqrt{\lambda_t},\,\sqrt{\tau}\bigr),
\label{eq:detection_prob}
\vspace{-0.02 in}\end{equation}
where $Q_{\chi_2^2}(\cdot,\cdot)$ denotes the complementary cumulative distribution function (CDF) of the noncentral Chi-squared distribution, and $Q_1(\cdot,\cdot)$ is the first-order Marcum $Q$-function. Equations \eqref{eq:lambda_t}-\eqref{eq:detection_prob} highlight that, for fixed radar parameters, detection performance is governed by the residual variance $\sigma_{\omega,\mathrm{r}}^{2(\varsigma)}$: lower $\sigma_{\omega,\mathrm{r}}^{2(\varsigma)}$ yields larger $\lambda_t$ and higher $P_{\mathrm{D},t}$ for a given $P_{\mathrm{FA}}$.

\section{Kullback--Leibler Divergence}

This section derives the KLD for all four operating configurations. The KLD serves as a unified performance measure for ISAC systems~\cite{Al-Jarrah2023}: in communication it is linked to the SER through pairwise ML detection, so that a larger KLD lowers the error probability, whereas in radar it is linked to detection performance through the GLRT framework and the miss-detection exponent given by Stein's lemma. This common distinguishability interpretation makes KLD a natural ISAC performance measure.
\label{r12c}
 For a pair of multivariate Gaussian distributed variables with mean vectors $\boldsymbol{\mu}_{m}$ and $\boldsymbol{\mu}_{n}$, and covariance matrices $\boldsymbol{\Sigma}_{m}$ and $\boldsymbol{\Sigma}_{n}$, the KLD is \cite{10.5555/1146355}
\begin{multline}
\mathrm{KLD}_{n\rightarrow m}=\frac{1}{2\ln 2}\left( \operatorname{tr}\left(
\boldsymbol{\Sigma}_{n}^{-1}\boldsymbol{\Sigma}_{m}\right) -2+\left( \boldsymbol{\mu}_{n}-\boldsymbol{\mu}_{m}\right)^{T} \right. \\
\left. \times\;\boldsymbol{\Sigma}_{n}^{-1} \left( \boldsymbol{\mu}_{n}-\boldsymbol{\mu}_{m}\right) +\ln \frac{\left\vert \boldsymbol{\Sigma}_{n}\right\vert }{\left\vert \boldsymbol{\Sigma}_{m}\right\vert }\right).\label{eq:KLD_general}
\end{multline}
 The four configurations are organised along two axes: the deployment scenario, SE or SH, sets the communication and radar array dimensions together with the effective pathloss, whereas the duplexity, HD or FD, affects only the interference-plus-noise statistics. Consequently, within a given deployment the desired-signal component is common to the two duplexities, and the entire configuration dependence of the expressions below is carried by the $(\varsigma)$-indexed variances $\sigma_{\omega,\mathrm{d},k}^{2(\varsigma)}$, $\sigma_{\omega,\mathrm{c}}^{2(\varsigma)}$, and $\sigma_{\omega,\mathrm{r}}^{2(\varsigma)}$ of Section~II. The SE deployment is derived first and in full; the SH deployment then follows by replacing the SE array dimensions $M_{\mathrm{c}},M_{\mathrm{r}}$ with the SH aperture $M$ and using the corresponding $(\varsigma)$ variances.\label{r39d}

\vspace{-0.05 in}
\subsection{Separated Deployment}
\subsubsection{Downlink}
Let $y_{k,l}^{(\mathrm{DL})}$ denote the $k$-th UE's entry of the SE DL signal in \eqref{eq:1} under HD or \eqref{eq:DL_SEFD} under FD, which after ZF precoding retains the desired symbol $s_{k,l}$ free of inter-user interference, while the radar-induced interference and noise, together with the inter-UE interference under FD, form the circularly-symmetric complex Gaussian term of variance $\sigma_{\omega,\mathrm{d},k}^{2(\varsigma)}$ in \eqref{eq:SE_DL_var}. Circular symmetry renders the real and imaginary parts independent and equal in variance, so the real representation $\mathbf{y}_{k} = [\Re\{y_{k,l}^{(\mathrm{DL})}\}, \Im\{y_{k,l}^{(\mathrm{DL})}\}]^T$ is a two-dimensional Gaussian with covariance $\sigma_{\omega,\mathrm{d},k}^{2(\varsigma)}\mathbf{I}_2$, identical under every transmitted symbol. For an arbitrary $M_d$-ary constellation $\{s_1,\ldots,s_{M_d}\}$ with $\mathbb{E}\{|s_n|^2\}=1$, the two hypotheses $s_{n}$ and $s_{m}$, $n\neq m$, have mean vectors $\boldsymbol{\mu}_{k,n} = \sqrt{P_{\mathrm{c},k}}\bar{d}_{\mathrm{c},k}^{-\eta/2}\alpha_{\mathrm{ZF}}[\Re\{s_{n}\}, \Im\{s_{n}\}]^T$ and $\boldsymbol{\mu}_{k,m} = \sqrt{P_{\mathrm{c},k}}\bar{d}_{\mathrm{c},k}^{-\eta/2}\alpha_{\mathrm{ZF}}[\Re\{s_{m}\}, \Im\{s_{m}\}]^T$, with effective average pathloss $\bar{d}_{\mathrm{c},k}^{-\eta/2} = \frac{1}{M_{\mathrm{c}}}\sum_{i=1}^{M_{\mathrm{c}}}d_{\mathrm{c},k,i}^{-\eta/2}$. Substituting these means into \eqref{eq:KLD_general} yields the pairwise KLD, conditioned on the symbol pair $(n,m)$ and on the channel realisation through $\alpha_{\mathrm{ZF}}$,
\vspace{-0.04 in}\begin{align}
\mathrm{KLD}_{k,n\rightarrow m}(\alpha_{\mathrm{ZF}}) &= \frac{1}{2\sigma_{\omega,\mathrm{d},k}^{2(\varsigma)}\ln2}|\boldsymbol{\mu}_{k,n} - \boldsymbol{\mu}_{k,m}|^2 \nonumber \\
&= \frac{P_{\mathrm{c},k}\bar{d}_{\mathrm{c},k}^{-\eta}\alpha_{\mathrm{ZF}}^2\,|s_{n} - s_{m}|^2}{2\sigma_{\omega,\mathrm{d},k}^{2(\varsigma)}\ln2}.
\vspace{-0.02 in}\label{eq:pairwise_kld}  \end{align}
For equiprobable signalling, averaging \eqref{eq:pairwise_kld} over the $M_d(M_d-1)$ dissimilar symbol pairs and over the ZF gain $\alpha_{\mathrm{ZF}}^2$, with $\lambda\triangleq\frac{1}{2}\sum_{n=1}^{M_d}\sum_{\substack{m=1\\m\neq n}}^{M_d}|s_{n} - s_{m}|^2$ set by the constellation geometry and $\bar{\alpha}_{\mathrm{ZF}}^2=\mathbb{E}\{\alpha_{\mathrm{ZF}}^2\}=2\sigma_H^2(M_{\mathrm{c}}-K_{\mathrm{D}}+1)$, gives the closed-form DL KLD
\vspace{-0.05 in}\begin{equation}
\mathrm{KLD}_{\mathrm{c},k}^{(\mathrm{DL},\varsigma)} = \frac{\lambda P_{\mathrm{c},k}\bar{\alpha}_{\mathrm{ZF}}^2\bar{d}_{\mathrm{c},k}^{-\eta}}{M_d(M_d-1)\sigma_{\omega,\mathrm{d},k}^{2(\varsigma)}\ln2}, \quad \forall k\in\mathcal K_{\mathrm D}.
\label{eq:KLD_SEHD_DL}
\vspace{-0.02 in}\end{equation} 
The DL KLD scales linearly with the transmit power $P_{\mathrm{c},k}$ and the ZF array gain $(M_{\mathrm{c}}-K_{\mathrm{D}}+1)$, and is inversely affected by radar-induced interference through $\sigma_{\omega,\mathrm{d},k}^{2(\varsigma)}$, whose FD value additionally contains the inter-UE interference scaling with the UL powers $\grave{P}_{\mathrm{c},j}$. This captures the fundamental communication--radar trade-off, since increasing $P_{\mathrm{r}}$ degrades DL performance.

\subsubsection{Uplink}
After IC, the CPU applies the ZF combiner $\mathbf{G}_{\mathrm{c}}$ to the received signal $\mathbf{y}_l^{(\mathrm{UL},\varsigma)}$ in \eqref{eq:UL_afterIC}, giving for the $k$-th UE
\vspace{-0.05 in}\begin{equation}
\!\!\tilde{u}_{k,l} \!= \![\mathbf{G}_{\mathrm{c}}\mathbf{y}_l^{(\mathrm{UL},\varsigma)}]_k \!\!=\! \!\sqrt{\grave{P}_{\mathrm{c},k}}\bar{\grave{d}}_{\mathrm{c},k}^{-\eta/2}u_{k,l} \!+\! [\mathbf{G}_{\mathrm{c}}(\boldsymbol{\omega}_{\mathrm{c}}^{(\varsigma)} \!\!+\! \mathbf{n}_l)]_k.
\label{eq:ul_sehd}
\vspace{-0.02 in}\end{equation}
Here $\bar{\grave{d}}_{\mathrm{c},k}^{-\eta/2}=\frac{1}{M_{\mathrm{c}}}\sum_{i=1}^{M_{\mathrm{c}}}\grave{d}_{\mathrm{c},i,k}^{-\eta/2}$ is the UL effective average pathloss. The ZF combiner eliminates inter-user interference while shaping the residual interference and noise; the aggregate interference-plus-noise after combining, a linear combination of the Gaussian residual $\boldsymbol{\omega}_{\mathrm{c}}^{(\varsigma)}$ and noise $\mathbf{n}_l$, is therefore complex Gaussian with variance $\sigma_{\xi,k}^{2(\varsigma)} = (\sigma_{\omega,\mathrm{c}}^{2(\varsigma)} + \sigma_{n}^{2})[(\grave{\mathbf{H}}^{H}\grave{\mathbf{H}})^{-1}]_{k,k}$, where $\sigma_{\omega,\mathrm{c}}^{2(\varsigma)}$ is given in \eqref{eq:SE_UL_var}. Following the DL derivation, the pairwise UL KLD is
\vspace{-0.05 in}\begin{equation}
\mathrm{KLD}_{k,n\rightarrow m}^{(\mathrm{UL})} = \frac{\grave{P}_{\mathrm{c},k}\bar{\grave{d}}_{\mathrm{c},k}^{-\eta}|s_n - s_m|^2}{2\sigma_{\xi,k}^{2(\varsigma)}\ln2}.
\vspace{-0.02 in}\end{equation}
Substituting $\sigma_{\xi,k}^{2(\varsigma)}$ and noting that, for i.i.d. $\mathcal{CN}(0,2\sigma_H^2)$ entries, $\grave{\mathbf{H}}^{H}\grave{\mathbf{H}}$ is complex Wishart with $M_{\mathrm{c}}$ degrees of freedom, so that $\mathbb{E}\{(\grave{\mathbf{H}}^{H}\grave{\mathbf{H}})^{-1}\} = \frac{\mathbf{I}_{K_{\mathrm{U}}}}{2\sigma_H^2(M_{\mathrm{c}} - K_{\mathrm{U}})}$ for $M_{\mathrm{c}} > K_{\mathrm{U}}$, the average UL KLD is
\vspace{-0.05 in}\begin{equation}
\!\!\!\!\mathrm{KLD}_{\mathrm{c},k}^{(\mathrm{UL},\varsigma)}\!=\!\! 
\frac{2\lambda\grave{P}_{\mathrm{c},k}\bar{\grave{d}}_{\mathrm{c},k}^{-\eta}\sigma_H^2(M_{\mathrm{c}} \!\!-\! K_{\mathrm{U}})}{M_d(M_d\!-\!\!1)(\sigma_{\omega,\mathrm{c}}^{2(\varsigma)} \!\!+\! \sigma_{n}^{2})\ln2}, \!\!\!\!\quad \forall k\in\mathcal K_{\mathrm U}.
\label{eq:KLD_SEHD_UL}
\vspace{-0.02 in}\end{equation}
Relative to DL, the UL KLD benefits from the ZF combining gain $(M_{\mathrm{c}} - K_{\mathrm{U}})$ but is degraded by the residual radar backscattering $\sigma_{\omega,\mathrm{c}}^{2(\varsigma)}$, which depends on the IC quality through the estimation-error variance $\sigma_{\mathrm{UH},\mathrm{err}_1}^{2}$. Under FD this variance additionally contains the imperfectly cancelled SI scaled by $\beta_{\mathrm{AP}}^2$, so that achieving $\beta_{\mathrm{AP}}\ll1$ through effective cancellation is critical for FD viability. The total average communication KLD, accounting for both directions, is
\vspace{-0.05 in}\begin{equation}
\mathrm{KLD}_{\mathrm{c}}^{(\varsigma)} = \frac{K_{\mathrm{D}}}{K}\sum_{k=1}^{K_{\mathrm{D}}}\mathrm{KLD}_{\mathrm{c},k}^{(\mathrm{DL},\varsigma)} + \frac{K_{\mathrm{U}}}{K}\sum_{j=1}^{K_{\mathrm{U}}}\mathrm{KLD}_{\mathrm{c},j}^{(\mathrm{UL},\varsigma)},
\label{eq:KLD_comm_total}
\vspace{-0.02 in}\end{equation}
which provides a unified communication measure balancing the DL and UL contributions according to the user counts $K_{\mathrm{D}}$ and $K_{\mathrm{U}}$.

 The average communication KLD governs the SER for any equiprobable constellation $\mathcal{C}=\{s_1,\ldots,s_{M_d}\}$ with $\mathbb{E}\{|s_n|^2\}=1$. Conditioned on $\alpha_{\mathrm{ZF}}$, the equal-covariance Gaussian model of \eqref{eq:KLD_general} makes the ML pairwise error probability a function of the conditional pairwise KLD in \eqref{eq:pairwise_kld},
\begin{equation}
P_{k,n\rightarrow m}(\alpha_{\mathrm{ZF}})=Q\!\left(\sqrt{\tfrac{\ln2}{2}\,\mathrm{KLD}_{k,n\rightarrow m}(\alpha_{\mathrm{ZF}})}\right),
\label{eq:pairwise_cond}
\end{equation}
where $Q(\cdot)$ is the standard Gaussian complementary CDF. As the pairwise KLD in \eqref{eq:pairwise_kld} is proportional to $|s_n-s_m|^2$, it relates to the per-configuration KLD through
\begin{equation}
\mathrm{KLD}_{k,n\rightarrow m}(\alpha_{\mathrm{ZF}})=\frac{\alpha_{\mathrm{ZF}}^2}{\bar{\alpha}_{\mathrm{ZF}}^2}\,\rho_{nm}\,\mathrm{KLD}_{\mathrm{c},k}^{(\mathrm{DL},\varsigma)},\quad \rho_{nm}\triangleq\frac{|s_n-s_m|^2}{\bar{d}_{\mathcal{C}}^2},
\label{eq:pair_factor}
\end{equation}
where $\bar{d}_{\mathcal{C}}^2=\tfrac{1}{M_d(M_d-1)}\sum_{i}\sum_{j\neq i}|s_i-s_j|^2$ is the mean squared distance, $\tfrac{1}{M_d(M_d-1)}\sum_{n}\sum_{m\neq n}\rho_{nm}=1$, and $\nu\triangleq M_{\mathrm{c}}-K_{\mathrm{D}}+1$, so that $\bar{\alpha}_{\mathrm{ZF}}^2=2\sigma_H^2\nu$. Substituting \eqref{eq:pair_factor} into \eqref{eq:pairwise_cond} gives $P_{k,n\rightarrow m}(\alpha_{\mathrm{ZF}})=Q\bigl(\alpha_{\mathrm{ZF}}\sqrt{a_{nm}/(2\sigma_H^2)}\bigr)$ with $a_{nm}=\tfrac{\ln2}{2\nu}\rho_{nm}\mathrm{KLD}_{\mathrm{c},k}^{(\mathrm{DL},\varsigma)}$, so deployment and duplexity affect the pairwise error probability through the per-configuration $\mathrm{KLD}_{\mathrm{c},k}^{(\mathrm{DL},\varsigma)}$. Since $\alpha_{\mathrm{ZF}}^2\sim\mathrm{Gamma}(\nu,2\sigma_H^2)$ under i.i.d. Rayleigh fading \cite[Sec.~8.3]{tse2005fundamentals}, averaging the conditional error over the fading yields the closed form \cite[Ch.~13, Eq. 13.4--15]{proakis2008digital},
\begin{equation}
\bar{P}_{k,n\rightarrow m}=\left(\frac{1-\mu_{nm}}{2}\right)^{\!\nu}\sum_{j=0}^{\nu-1}\binom{\nu-1+j}{j}\!\left(\frac{1+\mu_{nm}}{2}\right)^{\!j}\!,
\label{eq:pairwise_closed}
\end{equation}
where $\mu_{nm}=\sqrt{a_{nm}/(2+a_{nm})}$ and $\bar{P}_{k,n\rightarrow m}$ is the fading-averaged pairwise error. Averaging over the equiprobable symbol pairs, the SER admits the union upper bound \cite[Ch.~4, Eq. 4.2--70]{proakis2008digital},
\begin{equation}
P_{\mathrm{s},k}\le\frac{1}{M_d}\sum_{n=1}^{M_d}\sum_{\substack{m=1\\m\neq n}}^{M_d}\bar{P}_{k,n\rightarrow m},
\label{eq:union_ser}
\end{equation}
which expresses the SER as a function of the per-configuration KLD $\mathrm{KLD}_{\mathrm{c},k}^{(\mathrm{DL},\varsigma)}$ alone. For large $\mathrm{KLD}_{\mathrm{c},k}^{(\mathrm{DL},\varsigma)}$ the bound is dominated by the minimum-distance pairs, giving the nearest-neighbour approximation $P_{\mathrm{s},k}\approx\bar{N}_{\min}\bar{P}_{\min}$, where $\bar{P}_{\min}$ is \eqref{eq:pairwise_closed} evaluated at $\rho_{\min}=d_{\min}^2/\bar{d}_{\mathcal{C}}^2$, $d_{\min}=\min_{n\neq m}|s_n-s_m|$, and $\bar{N}_{\min}$ is the average number of nearest neighbours. The same relations hold for the UL using $\mathrm{KLD}_{\mathrm{c},k}^{(\mathrm{UL},\varsigma)}$ and the UL ZF gain $M_{\mathrm{c}}-K_{\mathrm{U}}$ in place of $\nu$.
\label{r12d}

\subsubsection{Radar System}
For radar, the CPU coherently combines the $L$ matched-filtered snapshots for the $t$-th target into the sufficient statistic $\bar{y}_{\mathrm{r},t|\mathcal{H}_q}=\operatorname{tr}(\mathbf{E}_{t,q})=\frac{1}{L}\sum_{l=1}^{L}\mathbf{w}_{\mathrm{r},t,l}^{H}\mathbf{y}_{\mathrm{r},t,l|\mathcal{H}_q}^{(\varsigma)}$ of \eqref{eq:sufficient_stat}. Applying \eqref{eq:KLD_general} per snapshot with the pathloss-weighted target response $\mathbf{g}_t\mathbf{g}_t^H$, where $\mathbf{g}_t=\mathbf{d}_{\mathrm{r},t}\odot\boldsymbol{\Theta}_t\mathbf{a}_t$ as in \eqref{eq:xi_asymptotic}, the matched-filter output has mean $\mu_{\mathcal{H}_1}=\alpha_t\sqrt{\tfrac{P_{\mathrm{r},t}}{M_{\mathrm{r}}}}\,\mathbf{g}_t^H\mathbf{R}_t\mathbf{g}_t$ and $\mu_{\mathcal{H}_0}=0$, with common per-snapshot residual variance per real dimension $\Sigma_{\mathcal{H}_0}=\Sigma_{\mathcal{H}_1}=\sigma_{\omega,\mathrm{r}}^{2(\varsigma)}\operatorname{tr}(\mathbf{R}_t)$; since $\Sigma_{\mathcal{H}_0}=\Sigma_{\mathcal{H}_1}$, the KLD is symmetric between hypotheses, $\mathrm{KLD}_{\mathrm{r},t}^{(\varsigma)}=\mathrm{KLD}_{t,\mathcal{H}_1\rightarrow\mathcal{H}_0}=\mathrm{KLD}_{t,\mathcal{H}_0\rightarrow\mathcal{H}_1}$, yielding
\vspace{-0.04 in}\begin{align}
\mathrm{KLD}_{\mathrm{r},t}^{(\varsigma)} &= \frac{1}{2\ln 2}\Sigma_{\mathcal{H}_1}^{-1}|\mu_{\mathcal{H}_1}|^2 \nonumber \\
&= \frac{|\alpha_t|^2P_{\mathrm{r},t}|\mathbf{g}_{t}^H\mathbf{R}_t\mathbf{g}_{t}|^2}{2\sigma_{\omega,\mathrm{r}}^{2(\varsigma)}M_{\mathrm{r}}\operatorname{tr}(\mathbf{R}_{t})\ln 2}, \!\!\!\!\quad\forall t\in\mathcal T,
\label{eq:KLD_SEHD_radar}
\vspace{-0.02 in}\end{align}
where $\mathbf{g}_t^H\mathbf{R}_t\mathbf{g}_t$ is the coherent beamforming gain toward bin $t$. The radar KLD is proportional to the target radar cross-section (RCS) $|\alpha_t|^2$, the radar power $P_{\mathrm{r},t}$, and the beamforming gain, and inversely proportional to the residual variance $\sigma_{\omega,\mathrm{r}}^{2(\varsigma)}$, establishing the radar--communication coupling whereby UL transmissions degrade radar detection unless effectively cancelled. The impact of residual synchronisation error admits a compact characterisation: when the residual per-AP phase-alignment offsets are modelled as i.i.d. zero-mean Gaussian phases $\theta_m$ with variance $\sigma_{\theta}^{2}$, the mean $\mu_{\mathcal{H}_1}$ becomes $\alpha_t\sqrt{\tfrac{P_{\mathrm{r},t}}{M_{\mathrm{r}}}}\,\mathbf{g}_t^{H}\mathbf{R}_t(\mathbf{g}_t\odot\mathbf{e}_{\theta})$, where $[\mathbf{e}_{\theta}]_m\triangleq\mathrm{e}^{j\theta_m}$ has expectation $\mathrm{e}^{-\sigma_{\theta}^{2}/2}$, so the radar KLD in \eqref{eq:KLD_SEHD_radar} is scaled by approximately $\mathrm{e}^{-\sigma_{\theta}^{2}}$ for small $\sigma_{\theta}^{2}$; since this factor is common to all four configurations, the configuration ordering is preserved. Under FD, continuous echo reception is gained, but the net benefit depends on whether this temporal advantage outweighs the three FD-specific contributions to $\sigma_{\omega,\mathrm{r}}^{2(\varsigma)}$: the DL-to-radar coupling, the radar SI scaled by $\beta_{\mathrm{R}}^2$, and the residual UL interference. The radar KLD is linked to detection performance through the GLRT framework in Section~III and through the following error-exponent interpretation. According to Stein's lemma, for the binary hypothesis testing problem $\{\mathcal{H}_0,\mathcal{H}_1\}$, at a fixed false-alarm probability, the miss probability of the optimal detector over $L$ snapshots decays exponentially at a rate equal to the per-snapshot KLD,
\vspace{-0.05 in}\begin{equation}
\lim_{L\to\infty}-\tfrac{1}{L}\log_2\!\bigl(1-P_{\mathrm{D},t}^{(\varsigma)}\bigr)=\mathrm{KLD}_{\mathrm{r},t}^{(\varsigma)},
\label{eq:stein_pd}
\vspace{-0.02 in}\end{equation}
so that $P_{\mathrm{D},t}^{(\varsigma)}\to1$ exponentially in $L\,\mathrm{KLD}_{\mathrm{r},t}^{(\varsigma)}$; maximising the radar KLD therefore maximises this error exponent; since the KLD and the noncentrality parameter of the test statistic, i.e., $\lambda_t$ as given in \eqref{eq:lambda_t}, both rise as $\sigma_{\omega,\mathrm{r}}^{2(\varsigma)}$ falls, the detection probability $P_{\mathrm{D},t}^{(\varsigma)}$ in \eqref{eq:detection_prob} improves alongside the radar KLD.
\label{r12e}

\vspace{-0.05 in}
\subsection{Shared Deployment}
SH deployment uses all $M$ APs for both functions, raising the ZF array gain to $(M-K_{\mathrm{D}}+1)$ but introducing clutter from the superimposed waveform. The derivations follow the SE case in \eqref{eq:KLD_SEHD_DL}--\eqref{eq:KLD_SEHD_radar}: since all $M$ APs now serve both functions, the separate communication and radar array sizes $M_{\mathrm{c}}$ and $M_{\mathrm{r}}$ are both replaced by the full array size $M$, the effective pathloss becomes $\bar{\dot{d}}_{\mathrm{c},k}^{-\eta/2} = \frac{1}{M}\sum_{i=1}^{M}\dot{d}_{\mathrm{c},k,i}^{-\eta/2}$, and the interference-plus-noise variances take their configuration-specific $(\varsigma)$ forms; as before, the duplexity enters only through these variances.
\subsubsection{Downlink}
With the composite-channel interference variance $\sigma_{\omega,\mathrm{d},k}^{2(\varsigma)}$ from \eqref{eq:SH_DL_var}, in which the clutter contributes through $\sigma_{F}^{2}$, the DL KLD is
\vspace{-0.05 in}\begin{equation}
\mathrm{KLD}_{\mathrm{c},k}^{(\mathrm{DL},\varsigma)} = \frac{\lambda P_{\mathrm{c},k}\bar{\alpha}_{\mathrm{ZF}}^2\bar{\dot{d}}_{\mathrm{c},k}^{-\eta}}{M_d(M_d-1)\sigma_{\omega,\mathrm{d},k}^{2(\varsigma)}\ln2}, \quad \forall k\in\mathcal K_{\mathrm D},
\label{eq:KLD_SHHD_DL}
\vspace{-0.02 in}\end{equation}
with $\bar{\alpha}_{\mathrm{ZF}}^2 = 2(\sigma_H^2+\sigma_F^2)(M-K_{\mathrm{D}}+1)$ reflecting the use of all $M$ antennas and the composite-channel variance. SH deployment thus increases the ZF gain but introduces the clutter variance $\sigma_{F}^{2}$ carried by $\sigma_{\omega,\mathrm{d},k}^{2(\varsigma)}$, creating a trade-off between array gain and interference.
\subsubsection{Uplink}
With $\sigma_{\omega,\mathrm{c}}^{2(\varsigma)}$ from \eqref{eq:SH_UL_var}, ZF combining gain $(M - K_{\mathrm{U}})$, and the UL effective average pathloss $\bar{\grave{d}}_{\mathrm{c},k}^{-\eta/2}=\frac{1}{M}\sum_{i=1}^{M}\grave{d}_{\mathrm{c},i,k}^{-\eta/2}$ taken over the full aperture, the UL KLD is
\vspace{-0.05 in}\begin{equation}
\!\!\!\!\mathrm{KLD}_{\mathrm{c},k}^{(\mathrm{UL},\varsigma)}\!=\!\! 
\frac{2\lambda\grave{P}_{\mathrm{c},k}\bar{\grave{d}}_{\mathrm{c},k}^{-\eta}\sigma_H^2(M \!-\! K_{\mathrm{U}})}{M_d(M_d\!-\!\!1)(\sigma_{\omega,\mathrm{c}}^{2(\varsigma)} \!\!+\! \sigma_{n}^{2})\!\ln\!2}, \!\!\quad \forall k\in\mathcal K_{\mathrm U}.
\label{eq:KLD_SHHD_UL}
\vspace{-0.02 in}\end{equation}
The increased count $M > M_{\mathrm{c}}$ raises the combining gain but also increases the residual interference through $\|\dot{\mathbf{D}}_{\mathrm{rc}}\|_{F}^{2}$ in $\sigma_{\omega,\mathrm{c}}^{2(\varsigma)}$; as in the SE case, the FD penalty is the SI scaled by $\beta_{\mathrm{AP}}^2$, for which $\beta_{\mathrm{AP}}\ll1$ is critical.
\subsubsection{Radar System}
With $\dot{\mathbf{g}}_t = \dot{\mathbf{d}}_{\mathrm{r},t} \odot \boldsymbol{\Theta}_t\dot{\mathbf{a}}_t$, transmit covariance $\dot{\mathbf{R}}_{t} = \frac{1}{L}\sum_{l=1}^{L}\dot{\mathbf{w}}_{\mathrm{r},t,l}\dot{\mathbf{w}}_{\mathrm{r},t,l}^{H}$, and $\sigma_{\omega,\mathrm{r}}^{2(\varsigma)}$ from \eqref{eq:SH_radar_var}, the radar KLD is
\vspace{-0.05 in}\begin{equation}
\mathrm{KLD}_{\mathrm{r},t}^{(\varsigma)} = \frac{|\alpha_t|^2P_{\mathrm{r},t}|\dot{\mathbf{g}}_{t}^H\dot{\mathbf{R}}_t\dot{\mathbf{g}}_{t}|^2}{2\sigma_{\omega,\mathrm{r}}^{2(\varsigma)}M\operatorname{tr}(\dot{\mathbf{R}}_{t})\ln 2}, \!\!\!\!\quad\forall t\in\mathcal T.
\label{eq:KLD_SHHD_radar}
\vspace{-0.02 in}\end{equation}
SH deployment gains the full $M$-antenna array but incurs SI from its own communication waveform, a coupling absent in SE deployment. Under FD this is compounded by the waveform SI scaled by $\beta_{\mathrm{R}}^{2}$ and by the residual clutter, which contributes to $\sigma_{\omega,\mathrm{r}}^{2(\varsigma)}$ through the AP-to-AP pathloss energy $\|\dot{\mathbf{D}}_{\mathrm{rc}}\|_{F}^{2}$ that reflects clutter accumulation across the array, while the target-reflected communication waveform contributes through the bin-$t$ pathloss energy $\|\dot{\mathbf{D}}_{\mathrm{r},t}\|_{F}^{2}$, common to the HD and FD cases. Achieving high SH-FD radar KLD therefore requires stringent SI suppression ($\beta\ll1$) alongside effective IC of all interference sources. Finally, the SER characterisation in \eqref{eq:pairwise_closed} and \eqref{eq:union_ser} and the detection-probability exponent \eqref{eq:stein_pd} depend on the deployment only through the KLDs they contain, and therefore apply verbatim to the SH deployment scenarios.
\section{Numerical Results}
This section presents simulation results evaluating the proposed CF-mMIMO ISAC framework under SE and SH deployments, comparing HD and FD operation. The system is deployed over an $800 \times 800$~m$^2$ coverage area, where APs are distributed using Poisson disk sampling with a minimum inter-AP separation of $100$~m, while UEs and radar targets are placed at random under the same minimum-separation constraint. Unless otherwise stated, the system serves $K_{\mathrm{D}}=2$ DL UEs and $K_{\mathrm{U}}=2$ UL UEs, while the radar subsystem detects $T=3$ targets. The SE deployment uses $M_{\mathrm{c}}=20$ communication APs and $M_{\mathrm{r}}=20$ radar APs, while SH deployment employs $M=40$ SH ISAC APs. $P_{\mathrm R}$ denotes the average desired received power after large-scale attenuation and $N_0\triangleq\sigma_n^2$ is the receiver-noise power.

Each Monte Carlo realisation consists of $L=100$ coherent snapshots, and all curves are obtained using $N_{\mathrm{MC}}=10^{6}$ channel realisations. Small-scale fading follows independent Rayleigh statistics, while large-scale attenuation follows a pathloss law with exponent $\eta=3$. The radar detector is configured with a false-alarm probability $P_{\mathrm{FA}}=10^{-4}$. Quadrature phase-shift keying (QPSK) is employed for HD and binary phase-shift keying (BPSK) for FD to equalise the effective data rate under the respective duplexity constraints, which is the operating point at which an error-rate comparison is meaningful; a common modulation order would instead compare the configurations at unequal rates. The closed-form analysis separates the two effects exactly: in both the KLD of \eqref{eq:KLD_SEHD_DL} and the SER of \eqref{eq:union_ser}, the constellation appears only through its geometry, namely the factor $\lambda$ in the KLD and the minimum distance $d_{\min}$ in the SER, while the deployment and duplexity are captured by the interference-plus-noise variance $\sigma_{\omega,\mathrm{d},k}^{2(\varsigma)}$ and its UL analogue. This variance is set by the transmit powers and the cancellation quality and is independent of the constellation, so the change in $\sigma_{\omega,\mathrm{d},k}^{2(\varsigma)}$ from HD to FD is the duplexity effect in isolation, obtained without a same-modulation comparison. Evaluated across all deployments, power splits, and $P_{\mathrm R}/N_0$ values, this variance is at most about $0.65$~dB higher under FD than under HD for the SH DL and essentially unchanged for the well-cancelled SE UL at $\sigma_{\mathrm{IC}}^2=10^{-6}$, rising only as cancellation quality degrades, consistent with the error floors at $\sigma_{\mathrm{IC}}^2$ or $\beta$ of $10^{-1}$. The change of duplexity is therefore close to interference-neutral when cancellation is effective, and the reported FD gains follow from the lower-order BPSK constellation that simultaneous transmission and reception permits at the matched rate: its squared minimum distance is twice that of QPSK, a $3$~dB advantage in symbol-error performance unavailable to HD, which cannot reduce its modulation order without halving its rate.\label{r13}

The total network-wide power budget is normalised and split among DL communication, radar probing, and UL transmission, denoted by $(P_{\mathrm{c}},P_{\mathrm{r}},P_{\mathrm{u}})$, with uniform per-user allocations $P_{\mathrm{c}}/K_{\mathrm{D}}$ and $P_{\mathrm{u}}/K_{\mathrm{U}}$. Performance is evaluated versus $P_{\mathrm{R}}/N_{0}$ and is reported using $\mathrm{KLD}_{\mathrm{c}}$ and SER for the communication subsystem, and $\mathrm{KLD}_{\mathrm{r}}$ and the detection probability $P_{\mathrm{D}}$ for the sensing subsystem.
This pairing makes the SER--KLD relation in \eqref{eq:union_ser} and the radar detection interpretation in \eqref{eq:detection_prob} and \eqref{eq:stein_pd} directly observable: across the figures the measured SER falls as $\mathrm{KLD}_{\mathrm{c}}$ increases and $P_{\mathrm{D}}\to1$ as $\mathrm{KLD}_{\mathrm{r}}$ grows.

To capture the practical limitations of FD operation, residual transmit--receive leakage is modelled through the leakage coefficient $\beta$ (or $\beta_{\mathrm{AP}}$ and $\beta_{\mathrm{R}}$ in SE deployment), while imperfect IC is characterised by the error variance $\sigma_{\mathrm{IC}}^{2}$. These are normalised residual-quality parameters, read in decibels as $20\log_{10}\beta$ and $10\log_{10}\sigma_{\mathrm{IC}}^{2}$: smaller values indicate stronger SI suppression and more accurate IC, respectively, and the values considered here span strong, moderate, and adverse cancellation levels.

The four configurations differ in implementation burden. SH deployment raises the processing dimension from the subsystem apertures $M_{\mathrm{c}}$ and $M_{\mathrm{r}}$ to the full aperture $M$. ZF precoding and combining invert a Gram of fixed order $K_{\mathrm{D}}$ or $K_{\mathrm{U}}$ and cost $\mathcal{O}(MK_{\mathrm{D}}^{2}+K_{\mathrm{D}}^{3})$ and $\mathcal{O}(MK_{\mathrm{U}}^{2}+K_{\mathrm{U}}^{3})$, linear in the aperture, whereas the $T$ GLRT statistics, each formed from the aperture-sized sufficient statistic in \eqref{eq:sufficient_stat}, together cost $\mathcal{O}(TM^{2})$, quadratic in the aperture; SE deployment replaces $M$ by $M_{\mathrm{c}}$ or $M_{\mathrm{r}}$. FD adds a dense $\mathcal{O}(M^{2})$ SI-cancellation stage absent under HD. For the present split $M=40$ and $M_{\mathrm{c}}=M_{\mathrm{r}}=20$, SH deployment approximately doubles the communication cost and quadruples the radar cost of SE deployment, while the SH-FD SI-cancellation stage costs about twice its SE-FD counterpart ($M^{2}$ vs $M_{\mathrm{c}}^{2}+M_{\mathrm{r}}^{2}$). SH-FD therefore attains the largest aperture and the strongest communication performance at the highest implementation burden, suiting deployments with sufficient fronthaul, CPU, and SI suppression, while SE or HD operation remains a lighter option within the same framework.
\label{r29b}
\label{r29}

\begin{figure}[!ht]
\centering
\vspace{-0.07in}
\includegraphics[width=3.3in]{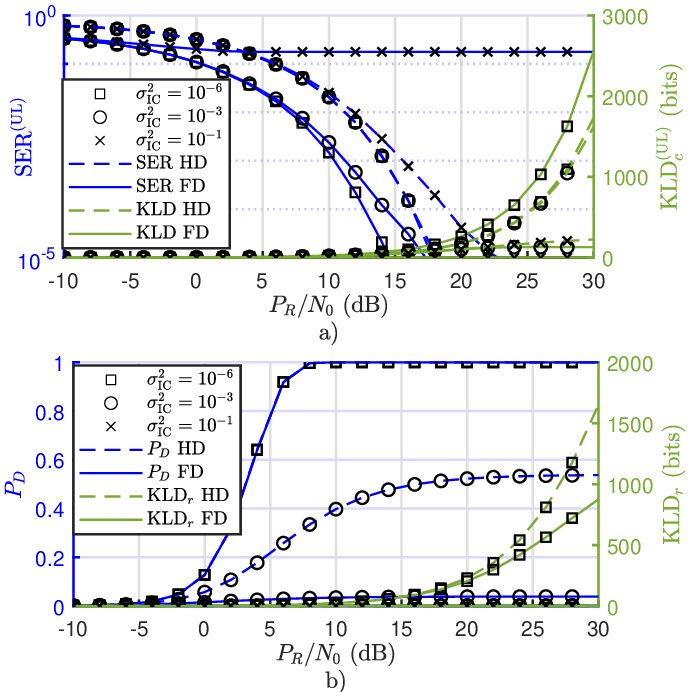} 
\vspace{-0.1in}
\caption{SE-ISAC performance under varying IC qualities. a) $\mathrm{KLD}_\mathrm{c}^{\mathrm{(UL)}}$ and $\mathrm{SER}^{(\mathrm{UL})}$ vs $P_\mathrm{R}/N_0$. b) $\mathrm{KLD}_\mathrm{r}$ and $P_\mathrm{D}$ vs $P_\mathrm{R}/N_0$.}
\label{fig:1}
\vspace{-0.09 in}
\end{figure}

Fig.~\ref{fig:1} presents SE-ISAC performance under varying IC qualities $\sigma_{\mathrm{IC}}^{2}\in\{10^{-6},\,10^{-3},\,10^{-1}\}$ with fixed $(P_{\mathrm c},P_{\mathrm r},P_{\mathrm u})=(0.8,0.1,0.1)$ and $\beta_{\mathrm{AP}}=\beta_{\mathrm{R}}=10^{-3}$. Fig.~\ref{fig:1}.a shows the UL communication performance. FD operation with high-quality IC demonstrates a significant advantage over HD. At $P_{\mathrm{R}}/N_{0}=10$~dB, the $\mathrm{SER}^{(\mathrm{UL})}$ values for HD and FD at $\sigma_{\mathrm{IC}}^{2}=10^{-6}$ are approximately $\{2.1\times10^{-2},\,1.5\times10^{-3}\}$, respectively, while the corresponding $\mathrm{KLD}_{\mathrm{c}}^{(\mathrm{UL})}$ values reach $\{17.35,\,26.01\}$~bits, demonstrating a $50\%$ improvement for FD. At $30$~dB the $\mathrm{KLD}_{\mathrm{c}}^{(\mathrm{UL})}$ reaches $\{1734.78,\,2563.49\}$~bits, a consistent $\sim48\%$ FD gain. For moderate IC quality ($\sigma_{\mathrm{IC}}^{2}=10^{-3}$), FD still achieves gains at lower $P_{\mathrm{R}}/N_0$ but experiences saturation at high $P_{\mathrm{R}}/N_{0}$, with $\mathrm{KLD}_{\mathrm{c}}^{(\mathrm{UL})}$ reaching only $129.81$~bits at $30$~dB due to residual interference. In contrast, with poor IC ($\sigma_{\mathrm{IC}}^{2}=10^{-1}$), FD exhibits a pronounced error floor at $\mathrm{SER}^{(\mathrm{UL})}\approx0.177$ and $\mathrm{KLD}_{\mathrm{c}}^{(\mathrm{UL})}\approx1.30$~bits, as residual interference dominates UL reception.

Fig.~\ref{fig:1}.b presents the radar sensing performance. At $P_{\mathrm{R}}/N_{0}=5$~dB, both HD and FD at $\sigma_{\mathrm{IC}}^{2}=10^{-6}$ achieve comparable detection probabilities $P_{\mathrm{D}}\approx\{0.781,\,0.779\}$, reaching $P_{\mathrm{D}}=1$ by $10$~dB. The corresponding $\mathrm{KLD}_{\mathrm{r}}$ values at $10$~dB are $\{23.94,\,23.63\}$~bits for HD and FD, demonstrating near-identical radar performance. By $P_{\mathrm{R}}/N_{0}=30$~dB, the $\mathrm{KLD}_{\mathrm{r}}$ reaches $\{1650.42,\,875.41\}$~bits, showing that while FD achieves excellent detection ($P_{\mathrm{D}}=1$), the $\mathrm{KLD}_{\mathrm{r}}$ is reduced compared to HD due to additional interference from simultaneous communication. For moderate IC quality ($\sigma_{\mathrm{IC}}^{2}=10^{-3}$), HD achieves $P_{\mathrm{D}}\approx0.523$ while FD remains at $P_{\mathrm{D}}\approx0.039$ at $20$~dB, with corresponding $\mathrm{KLD}_{\mathrm{r}}$ values of $\{5.15,\,1.37\}$~bits. These results establish that FD provides substantial $\mathrm{KLD}_{\mathrm{c}}^{(\mathrm{UL})}$ gains ($\sim48\%$) when $\sigma_{\mathrm{IC}}^{2}\leq10^{-6}$ while maintaining excellent radar detection, validating the FD-ISAC approach.

\begin{figure}[!ht]
\centering
\vspace{-0.07in}
\includegraphics[width=3.3in]{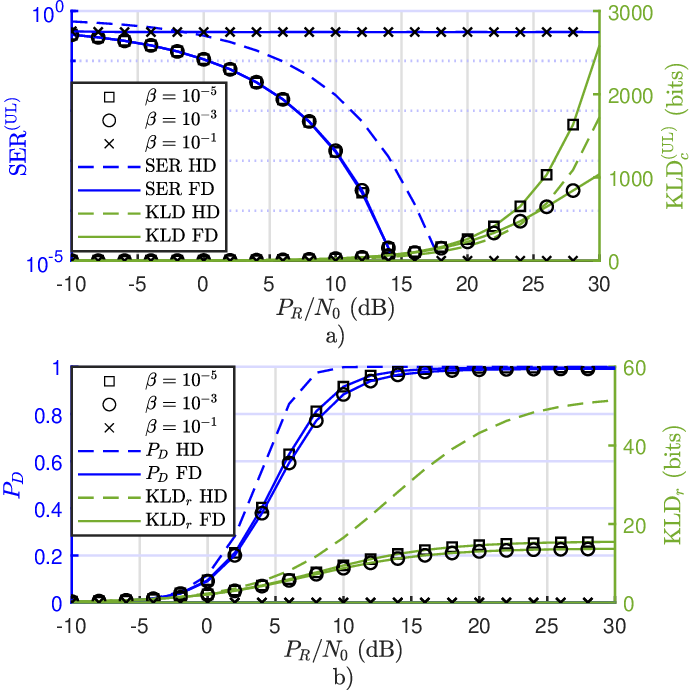} 
\vspace{-0.1in}
\caption{SE-ISAC performance under varying SI leakage levels. a) $\mathrm{KLD}_\mathrm{c}^{\mathrm{(UL)}}$ and $\mathrm{SER}^{(\mathrm{UL})}$ vs $P_\mathrm{R}/N_0$. b) $\mathrm{KLD}_\mathrm{r}$ and $P_\mathrm{D}$ vs $P_\mathrm{R}/N_0$.}
\label{fig:2}
\vspace{-0.12 in}
\end{figure}
Fig.~\ref{fig:2} examines SE-ISAC performance under varying SI leakage $\beta\in\{10^{-5},\,10^{-3},\,10^{-1}\}$, quantifying residual transmit--receive coupling after cancellation, with fixed $(P_{\mathrm c},P_{\mathrm r},P_{\mathrm u})=(0.8,0.1,0.1)$ and $\sigma_{\mathrm{IC}}^{2}=10^{-4}$. Fig.~\ref{fig:2}.a shows UL communication performance, where FD advantage is evident: at $P_{\mathrm{R}}/N_{0}=10$~dB, the $\mathrm{SER}^{(\mathrm{UL})}$ values are $\{2.1\times10^{-2},\,1.5\times10^{-3},\,1.6\times10^{-3}\}$ for HD, FD with $\beta=10^{-5}$, and FD with $\beta=10^{-3}$, respectively, an order-of-magnitude lower SER for FD. The corresponding $\mathrm{KLD}_{\mathrm{c}}^{(\mathrm{UL})}$ values are $\{17.36,\,26.00,\,25.62\}$~bits, a $\sim50\%$ FD gain. At $30$~dB the $\mathrm{KLD}_{\mathrm{c}}^{(\mathrm{UL})}$ reaches $\{1722.59,\,2584.76,\,1037.92\}$~bits for HD, $\beta=10^{-5}$, and $\beta=10^{-3}$. In contrast, large leakage ($\beta=10^{-1}$) results in a persistent error floor at $\mathrm{SER}^{(\mathrm{UL})}\approx0.378$ and $\mathrm{KLD}_{\mathrm{c}}^{(\mathrm{UL})}\approx0.13$~bits, rendering FD impractical.

Fig.~\ref{fig:2}.b presents the radar sensing performance. At $P_{\mathrm{R}}/N_{0}=10$~dB, the $P_{\mathrm{D}}$ values are $\{0.999,\,0.916,\,0.883\}$ for HD, FD with $\beta=10^{-5}$, and $\beta=10^{-3}$, respectively. By $20$~dB, FD with $\beta\leq10^{-3}$ achieves $P_{\mathrm{D}}\approx\{0.995,\,0.988\}$, closely approaching HD, while at $30$~dB the $\mathrm{KLD}_{\mathrm{r}}$ values are $\{51.50,\,15.46,\,13.69\}$~bits for HD, $\beta=10^{-5}$, and $\beta=10^{-3}$. For $\beta=10^{-1}$, detection probability collapses to $P_{\mathrm{D}}\approx0.0002$ with negligible $\mathrm{KLD}_{\mathrm{r}}\approx0.012$~bits. These results establish deployment guidelines: FD achieves $\sim50\%$ higher $\mathrm{KLD}_{\mathrm{c}}^{(\mathrm{UL})}$ while maintaining good radar detection when $\beta\leq10^{-3}$.

\begin{figure}[!ht]
\centering
\vspace{-0.07in}
\includegraphics[width=3.3in]{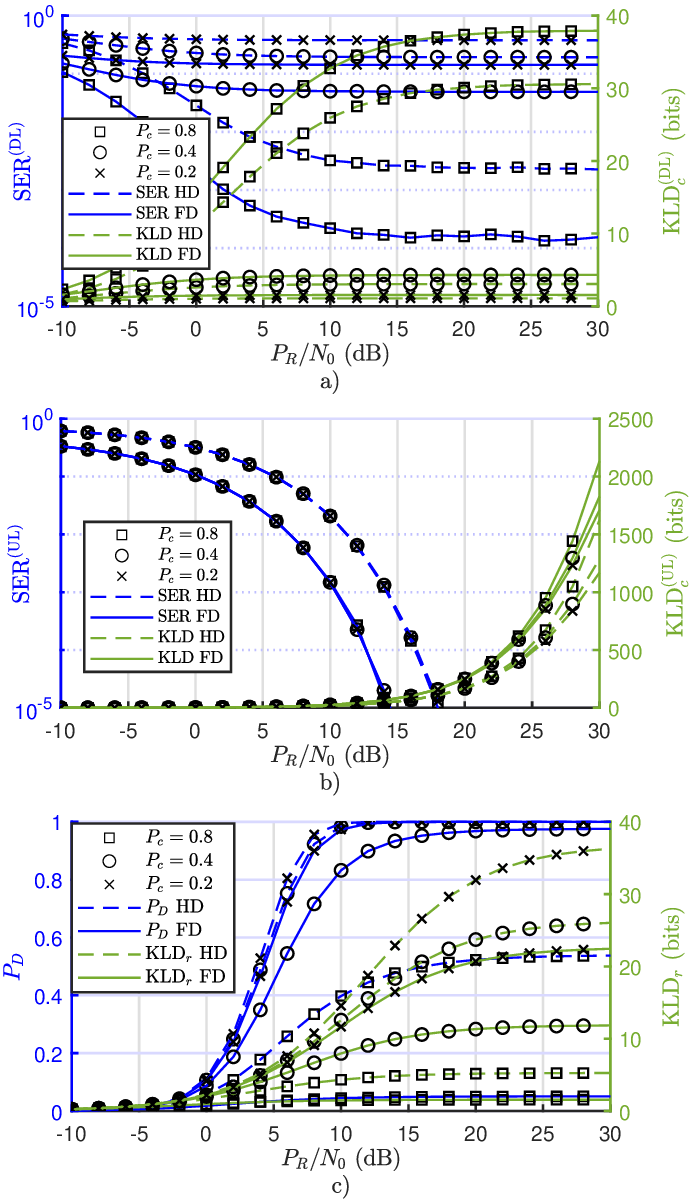} 
\vspace{-0.1in}
\caption{SE-ISAC performance under varying power allocation. a) $\mathrm{KLD}_\mathrm{c}^{\mathrm{(DL)}}$ and $\mathrm{SER}^{(\mathrm{DL})}$ vs $P_\mathrm{R}/N_0$. b) $\mathrm{KLD}_\mathrm{c}^{\mathrm{(UL)}}$ and $\mathrm{SER}^{(\mathrm{UL})}$ vs $P_\mathrm{R}/N_0$. c) $\mathrm{KLD}_\mathrm{r}$ and $P_\mathrm{D}$ vs $P_\mathrm{R}/N_0$.}
\label{fig:3}
\vspace{-0.14 in}
\end{figure}
Fig.~\ref{fig:3} analyses SE-ISAC performance under power allocations $P_{\mathrm{c}}\in\{0.8,\,0.4,\,0.2\}$ (i.e., $P_{\mathrm{r}}\in\{0.1,\,0.5,\,0.7\}$) with $P_{\mathrm{u}}=0.1$. The FD case uses $\beta_{\mathrm{AP}}=\beta_{\mathrm{R}}=10^{-4}$ and an IC-error variance $\sigma_{\mathrm{IC}}^{2}=10^{-3}$. Fig.~\ref{fig:3}.a shows the DL communication performance. FD consistently outperforms HD across all power allocations. At $P_{\mathrm{R}}/N_{0}=10$~dB with $P_{\mathrm{c}}=0.8$, the $\mathrm{SER}^{(\mathrm{DL})}$ values are $\{3.2\times10^{-3},\,2.2\times10^{-4}\}$ for HD and FD, with corresponding $\mathrm{KLD}_{\mathrm{c}}^{(\mathrm{DL})}$ of $\{25.94,\,32.74\}$~bits. At $30$~dB the $\mathrm{KLD}_{\mathrm{c}}^{(\mathrm{DL})}$ reaches $\{30.60,\,37.92\}$~bits, a consistent $\sim24\text{--}26\%$ FD gain. For lower $P_{\mathrm{c}}$ allocations, FD maintains its advantage: at $P_{\mathrm{c}}=0.4$ and $30$~dB, HD achieves $3.06$~bits while FD reaches $4.31$~bits.

Fig.~\ref{fig:3}.b presents the UL communication results. At $P_{\mathrm{R}}/N_{0}=10$~dB, the $\mathrm{SER}^{(\mathrm{UL})}$ values for HD and FD with $P_{\mathrm{c}}=0.8$ are $\{2.1\times10^{-2},\,1.5\times10^{-3}\}$. The $\mathrm{KLD}_{\mathrm{c}}^{(\mathrm{UL})}$ values at $30$~dB are $\{1622.41,\,2132.64\}$, $\{1287.00,\,1827.27\}$, and $\{1168.27,\,1709.02\}$~bits for $P_{\mathrm{c}}=0.8$, $0.4$, and $0.2$, respectively, a consistent $\sim32$--$46\%$ FD gain across all power allocations.

Fig.~\ref{fig:3}.c illustrates the radar sensing performance, where the power allocation impact is most pronounced. At $P_{\mathrm{R}}/N_{0}=10$~dB with $P_{\mathrm{c}}=0.2$ (hence $P_{\mathrm{r}}=0.7$), the $P_{\mathrm{D}}$ values are $\{0.995,\,0.974\}$ for HD and FD. By $15$~dB, HD achieves $P_{\mathrm{D}}\approx1.0$ while FD reaches $0.999$. For $P_{\mathrm{c}}=0.8$ (limited radar power), the $P_{\mathrm{D}}$ remains at $\{0.523,\,0.050\}$ even at $20$~dB. The corresponding $\mathrm{KLD}_{\mathrm{r}}$ values at $20$~dB are $\{5.15,\,1.54\}$~bits for $P_{\mathrm{c}}=0.8$, $\{23.72,\,11.33\}$~bits for $P_{\mathrm{c}}=0.4$, and $\{31.94,\,20.70\}$~bits for $P_{\mathrm{c}}=0.2$. This confirms the communication--sensing trade-off: FD sensing remains competitive when sufficient power is allocated to radar.

\begin{figure}[!ht]
\centering
\vspace{-0.07in}
\includegraphics[width=3.3in]{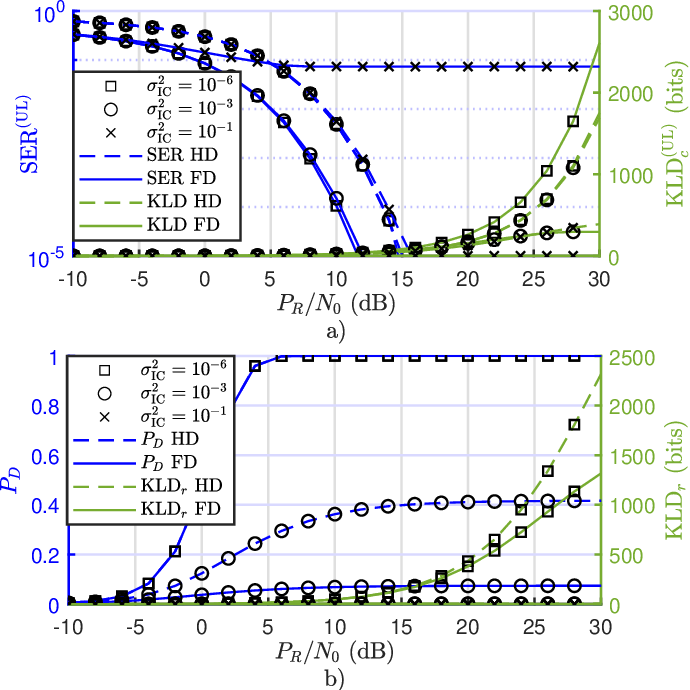} 
\vspace{-0.1in}
\caption{SH-ISAC performance under varying IC qualities. a) $\mathrm{KLD}_\mathrm{c}^{\mathrm{(UL)}}$ and $\mathrm{SER}^{(\mathrm{UL})}$ vs $P_\mathrm{R}/N_0$. b) $\mathrm{KLD}_\mathrm{r}$ and $P_\mathrm{D}$ vs $P_\mathrm{R}/N_0$.}
\label{fig:4}
\vspace{-0.12 in}
\end{figure}

Fig.~\ref{fig:4} reports SH-ISAC performance under varying IC quality with the power split and leakage of Fig.~\ref{fig:1}. Although SH deployment forgoes the subsystem isolation of SE (all $M=40$ APs support both functions), significant FD advantages remain. Fig.~\ref{fig:4}.a shows the UL communication performance. At $P_{\mathrm{R}}/N_{0}=10$~dB with $\sigma_{\mathrm{IC}}^{2}=10^{-6}$, the $\mathrm{SER}^{(\mathrm{UL})}$ values are $\{4.9\times10^{-3},\,1.1\times10^{-4}\}$ for HD and FD, representing a $\sim45\times$ improvement for FD. The corresponding $\mathrm{KLD}_{\mathrm{c}}^{(\mathrm{UL})}$ values are $\{17.51,\,26.26\}$~bits, showing a $50\%$ FD advantage. By $P_{\mathrm{R}}/N_{0}=15$~dB, the $\mathrm{SER}^{(\mathrm{UL})}$ drops to $\{2.7\times10^{-5},\,2.5\times10^{-7}\}$, and at $30$~dB the $\mathrm{KLD}_{\mathrm{c}}^{(\mathrm{UL})}$ reaches $\{1750.44,\,2607.47\}$~bits, a consistent $\sim49\%$ FD gain. For moderate IC quality ($\sigma_{\mathrm{IC}}^{2}=10^{-3}$), FD achieves $\mathrm{KLD}_{\mathrm{c}}^{(\mathrm{UL})}\approx295.84$~bits at $30$~dB, while poor IC ($\sigma_{\mathrm{IC}}^{2}=10^{-1}$) results in an error floor at $\mathrm{SER}^{(\mathrm{UL})}\approx0.073$ and $\mathrm{KLD}_{\mathrm{c}}^{(\mathrm{UL})}\approx2.96$~bits.

Fig.~\ref{fig:4}.b presents the SH radar sensing performance. The larger effective aperture ($M=40$) enables faster detection convergence. At $P_{\mathrm{R}}/N_{0}=5$~dB with $\sigma_{\mathrm{IC}}^{2}=10^{-6}$, both HD and FD achieve $P_{\mathrm{D}}\approx\{0.979,\,0.979\}$, reaching $P_{\mathrm{D}}=1$ by $10$~dB. The corresponding $\mathrm{KLD}_{\mathrm{r}}$ values at $10$~dB are $\{47.58,\,46.85\}$~bits, substantially exceeding the SE deployment values due to the larger array gain. At $20$~dB, HD achieves $\mathrm{KLD}_{\mathrm{r}}\approx434.18$~bits while FD reaches $380.08$~bits. For moderate IC quality ($\sigma_{\mathrm{IC}}^{2}=10^{-3}$), HD achieves $P_{\mathrm{D}}\approx0.411$ while FD remains at $P_{\mathrm{D}}\approx0.074$ at $20$~dB, with $\mathrm{KLD}_{\mathrm{r}}$ of $\{4.43,\,1.81\}$~bits. Even under SH deployment, where every AP serves both subsystems, FD achieves $\sim49\%$ communication gains while maintaining excellent radar detection when $\sigma_{\mathrm{IC}}^{2}\leq10^{-6}$.
\begin{figure}[!ht]
\centering
\vspace{-0.07in}
\includegraphics[width=3.3in]{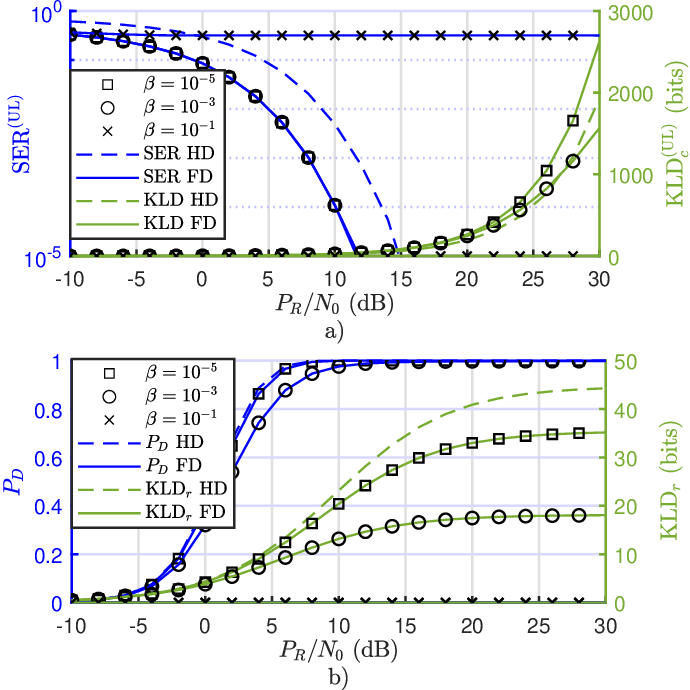} 
\vspace{-0.1in}
\caption{SH-ISAC performance under varying SI leakage levels. a) $\mathrm{KLD}_\mathrm{c}^{\mathrm{(UL)}}$ and $\mathrm{SER}^{(\mathrm{UL})}$ vs $P_\mathrm{R}/N_0$. b) $\mathrm{KLD}_\mathrm{r}$ and $P_\mathrm{D}$ vs $P_\mathrm{R}/N_0$.}
\label{fig:5}
\vspace{-0.08 in}
\end{figure}

Fig.~\ref{fig:5} evaluates SH-ISAC sensitivity to SI leakage $\beta$ with $(P_{\mathrm c},P_{\mathrm r},P_{\mathrm u})=(0.8,0.1,0.1)$ and $\sigma_{\mathrm{IC}}^{2}=10^{-4}$. Fig.~\ref{fig:5}.a shows that FD maintains substantial advantages for low leakage. At $P_{\mathrm{R}}/N_{0}=10$~dB, the $\mathrm{SER}^{(\mathrm{UL})}$ values are $\{4.9\times10^{-3},\,1.1\times10^{-4},\,1.1\times10^{-4}\}$ for HD, FD with $\beta=10^{-5}$, and $\beta=10^{-3}$, respectively. The corresponding $\mathrm{KLD}_{\mathrm{c}}^{(\mathrm{UL})}$ values are $\{19.24,\,26.23,\,26.07\}$~bits, showing $\sim36\%$ improvement for FD. By $15$~dB, FD achieves $\mathrm{SER}^{(\mathrm{UL})}<10^{-6}$ at both leakage levels, and at $30$~dB the $\mathrm{KLD}_{\mathrm{c}}^{(\mathrm{UL})}$ reaches $\{1924.75,\,2624.13,\,1566.20\}$~bits for HD, $\beta=10^{-5}$, and $\beta=10^{-3}$. In contrast, large leakage ($\beta=10^{-1}$) results in a persistent error floor at $\mathrm{SER}^{(\mathrm{UL})}\approx0.317$ and $\mathrm{KLD}_{\mathrm{c}}^{(\mathrm{UL})}\approx0.30$~bits.

Fig.~\ref{fig:5}.b demonstrates the radar sensitivity to SI leakage. At $P_{\mathrm{R}}/N_{0}=10$~dB, the $P_{\mathrm{D}}$ values are $\{0.9999,\,0.9994,\,0.9758\}$ for HD, FD with $\beta=10^{-5}$, and $\beta=10^{-3}$. By $20$~dB, FD with $\beta\leq10^{-3}$ achieves $P_{\mathrm{D}}\approx\{1.0,\,0.997\}$, essentially matching HD. At $30$~dB, the $\mathrm{KLD}_{\mathrm{r}}$ values are $\{44.28,\,35.15,\,18.06\}$~bits for HD, $\beta=10^{-5}$, and $\beta=10^{-3}$. For $\beta=10^{-1}$, detection probability collapses to $P_{\mathrm{D}}\approx0.00015$ with negligible $\mathrm{KLD}_{\mathrm{r}}\approx0.004$~bits. FD provides substantial communication gains while maintaining competitive radar performance when $\beta\leq10^{-3}$.

\begin{figure}[!ht]
\centering
\vspace{-0.07in}
\includegraphics[width=3.3in]{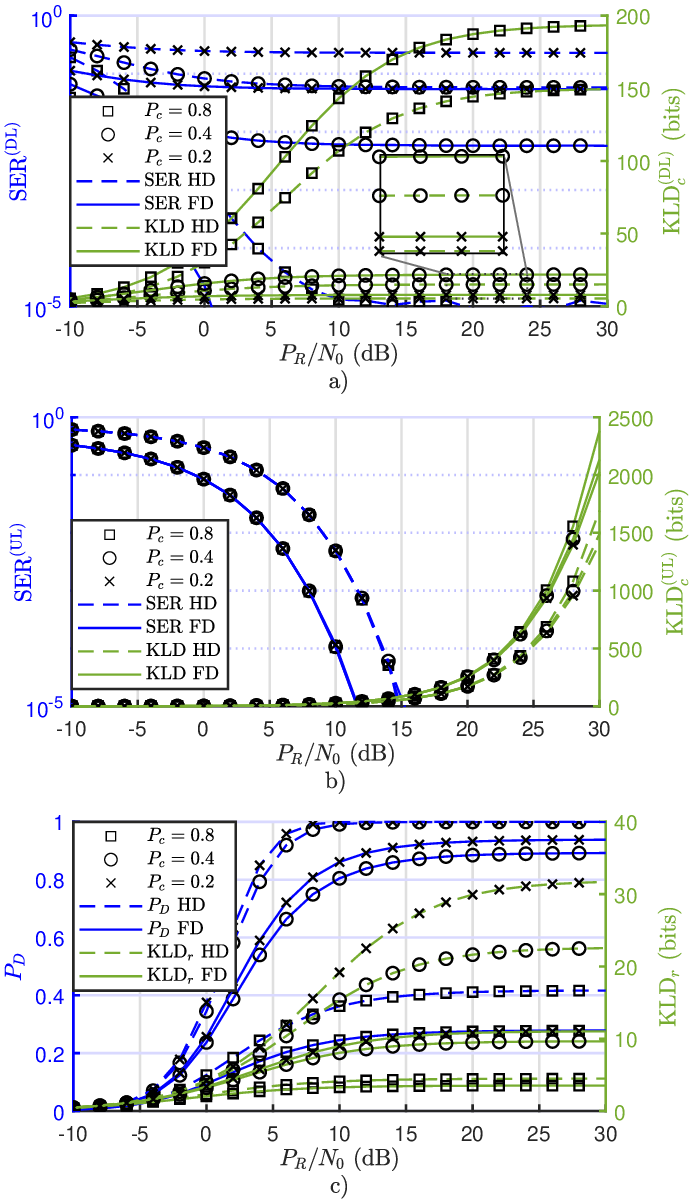} 
\vspace{-0.1in}
\caption{ SH-ISAC performance under varying power allocation. a) $\mathrm{KLD}_\mathrm{c}^{\mathrm{(DL)}}$ and $\mathrm{SER}^{(\mathrm{DL})}$ vs $P_\mathrm{R}/N_0$. b) $\mathrm{KLD}_\mathrm{c}^{\mathrm{(UL)}}$ and $\mathrm{SER}^{(\mathrm{UL})}$ vs $P_\mathrm{R}/N_0$. c) $\mathrm{KLD}_\mathrm{r}$ and $P_\mathrm{D}$ vs $P_\mathrm{R}/N_0$.}
\label{fig:6}\label{r311}
\vspace{-0.12 in}
\end{figure}

Fig.~\ref{fig:6} presents SH-ISAC performance under different power allocations. The FD case uses $\beta_{\mathrm{AP}}=\beta_{\mathrm{R}}=10^{-4}$ and an IC-error variance $\sigma_{\mathrm{IC}}^{2}=10^{-3}$. Fig.~\ref{fig:6}.a shows significant DL performance advantages for FD. At $P_{\mathrm{R}}/N_{0}=10$~dB with $P_{\mathrm{c}}=0.8$, the $\mathrm{SER}^{(\mathrm{DL})}$ values are $\{1.3\times10^{-5},\,{<}10^{-6}\}$ for HD and FD, with corresponding $\mathrm{KLD}_{\mathrm{c}}^{(\mathrm{DL})}$ of $\{106.96,\,143.08\}$~bits, representing a $34\%$ FD advantage. By $20$~dB, the $\mathrm{KLD}_{\mathrm{c}}^{(\mathrm{DL})}$ reaches $\{144.22,\,187.26\}$~bits, and at $30$~dB achieves $\{149.44,\,193.29\}$~bits, demonstrating consistent $\sim29\%$ improvement. For $P_{\mathrm{c}}=0.4$, FD achieves $\mathrm{KLD}_{\mathrm{c}}^{(\mathrm{DL})}\approx21.76$~bits versus HD's $15.00$~bits at $30$~dB, showing a $45\%$ improvement.

Fig.~\ref{fig:6}.b presents the UL communication performance. At $P_{\mathrm{R}}/N_{0}=10$~dB, the $\mathrm{SER}^{(\mathrm{UL})}$ values for HD and FD with $P_{\mathrm{c}}=0.8$ are $\{4.8\times10^{-3},\,1.0\times10^{-4}\}$, confirming FD's superior performance. The $\mathrm{KLD}_{\mathrm{c}}^{(\mathrm{UL})}$ values at $20$~dB are $\{174.44,\,259.82\}$~bits for $P_{\mathrm{c}}=0.8$, $\{172.02,\,256.47\}$~bits for $P_{\mathrm{c}}=0.4$, and $\{170.80,\,254.88\}$~bits for $P_{\mathrm{c}}=0.2$. At $30$~dB, these increase to $\{1691.58,\,2381.49\}$, $\{1490.51,\,2132.93\}$, and $\{1407.21,\,2027.93\}$~bits, showing consistent $\sim41$--$44\%$ FD improvement across all power allocations.

Fig.~\ref{fig:6}.c demonstrates the radar sensing trade-off. At $P_{\mathrm{R}}/N_{0}=10$~dB with $P_{\mathrm{c}}=0.2$, the $P_{\mathrm{D}}$ values are $\{0.999,\,0.861\}$ for HD and FD. By $15$~dB, HD achieves $P_{\mathrm{D}}\approx1.0$ while FD reaches $0.916$. For $P_{\mathrm{c}}=0.8$ (limited radar power), the $P_{\mathrm{D}}$ remains at $\{0.411,\,0.275\}$ even at $20$~dB. The corresponding $\mathrm{KLD}_{\mathrm{r}}$ values at $20$~dB are $\{4.43,\,3.48\}$~bits for $P_{\mathrm{c}}=0.8$, $\{21.62,\,9.47\}$~bits for $P_{\mathrm{c}}=0.4$, and $\{29.92,\,10.79\}$~bits for $P_{\mathrm{c}}=0.2$. Compared to SE in Fig.~\ref{fig:3}.c, SH demonstrates faster detection convergence due to a larger aperture $M=40$, though FD shows greater $\mathrm{KLD}_{\mathrm{r}}$ reduction due to the communication-induced interference at the shared aperture.

The impact of imperfect CSI and sensing estimation on the four configurations
is now examined with the receivers left unchanged, using
$(P_{\mathrm c},P_{\mathrm r},P_{\mathrm u})=(0.8,0.1,0.1)$,
$\sigma_{\mathrm{IC}}^{2}=10^{-3}$ and $\beta=10^{-4}$, QPSK and BPSK signalling
for HD and FD, and KLD in bits; each cell of Table~\ref{tab:robustness} lists
$P_{\mathrm{R}}/N_0\in\{10,20,30\}$~dB. For communication, the ZF precoder and combiner are reconstructed from the channel estimates
$\widehat{\mathbf{H}}=\mathbf{H}+\Delta\mathbf{H}$ and
$\widehat{\grave{\mathbf{H}}}=\grave{\mathbf{H}}+\Delta\grave{\mathbf{H}}$, whose
errors $\Delta\mathbf{H},\Delta\grave{\mathbf{H}}$ have i.i.d. $\mathcal{CN}$
entries at a normalised level
$\epsilon_{\mathrm c}=\mathbb{E}\{\|\Delta\mathbf{H}\|_F^{2}\}/\mathbb{E}\{\|\mathbf{H}\|_F^{2}\}
=\mathbb{E}\{\|\Delta\grave{\mathbf{H}}\|_F^{2}\}/\mathbb{E}\{\|\grave{\mathbf{H}}\|_F^{2}\}$, while the true channels govern propagation. The communication KLD grows with $P_{\mathrm{R}}/N_0$ for every configuration, and the comparative trends persist: SH deployment
attains a far larger DL KLD through its wider ZF aperture ($M=40$ versus
$M_{\mathrm c}=20$), the UL KLD is comparable across deployments, and FD exceeds HD in both directions. At $\epsilon_{\mathrm c}=10^{-1}$ the DL KLD falls by about $15\%$ and the UL by $30$--$40\%$, with the SER rising accordingly; the DL SER is interference-limited and therefore floors as $P_{\mathrm{R}}/N_0$ increases, whereas the UL SER is noise-limited and drops below the $10^{-6}$ Monte~Carlo resolution by $20$~dB.

For sensing, the GLRT is evaluated with the estimates the receiver holds, an
imperfect reflection coefficient $\hat{\alpha}_t=\alpha_t(1+\delta_{\alpha,t})$
and an imperfect response vector
$\hat{\mathbf{g}}_t^{(\varsigma)}=\mathbf{g}_t^{(\varsigma)}+\Delta\mathbf{g}_t^{(\varsigma)}$,
where $\Delta\mathbf{g}_t^{(\varsigma)}$ aggregates AP-position and range/bin
uncertainty and $\delta_{\alpha,t}$ is the RCS estimation error, both at a
normalised level $\epsilon_{\mathrm r}=\mathbb{E}\{|\delta_{\alpha,t}|^{2}\}
=\mathbb{E}\{\|\Delta\mathbf{g}_t^{(\varsigma)}\|^{2}\}/\mathbb{E}\{\|\mathbf{g}_t^{(\varsigma)}\|^{2}\}$,
with $\delta_{\alpha,t}$ scaled so that
$\mathbb{E}\{|\hat{\alpha}_t|^{2}\}=|\alpha_t|^{2}$, so that
$\epsilon_{\mathrm r}$ measures estimation accuracy rather than an average RCS
bias. This is the rank-one target model with a per-AP signature, set by the
AP--target geometry, and an unknown reflectivity, both recovered in distributed
ISAC by joint detection and channel estimation rather than known exactly
\cite{10557638}; as the detector matches the estimate it holds, the coherent
gain $\mathbf{g}_t^{H}\mathbf{R}_t\mathbf{g}_t$ in \eqref{eq:lambda_t},
\eqref{eq:KLD_SEHD_radar} and \eqref{eq:KLD_SHHD_radar} is replaced by the
matched-to-estimate gain
\begin{equation}
\frac{\bigl|(\hat{\mathbf{g}}_t^{(\varsigma)})^{H}\mathbf{R}_t\mathbf{g}_t^{(\varsigma)}\bigr|^{2}}
{(\hat{\mathbf{g}}_t^{(\varsigma)})^{H}\mathbf{R}_t\hat{\mathbf{g}}_t^{(\varsigma)}}
\;\le\;\mathbf{g}_t^{H}\mathbf{R}_t\mathbf{g}_t,
\label{eq:mismatch_loss}
\end{equation}
and $|\alpha_t|^{2}$ by $|\hat{\alpha}_t|^{2}$. HD retains a higher
radar KLD and detection probability than FD, with SE-HD highest; at this moderate-IC, residual-limited operating point, the radar KLD is of the order of a few bits and the detection probability is moderate, so a lower $\sigma_{\mathrm{IC}}^{2}$ would raise both while preserving the ordering. At $\epsilon_{\mathrm r}=10^{-1}$ the radar KLD falls by about $9\%$ with a commensurate detection-probability reduction.

At the realistic level $\epsilon_{\mathrm c}=\epsilon_{\mathrm r}=10^{-3}$ all measures move by under $1\%$, and across $P_{\mathrm{R}}/N_0$ and both impairments the ordering between deployments and duplexities are preserved, so the comparative conclusions are robust to imperfect channel and target knowledge, degrading gracefully rather than reordering. The two impairments act independently, as the radar detector does not use the communication CSI and the communication receivers do not use the target response, with $\epsilon_{\mathrm c}=\epsilon_{\mathrm r}=0$ recovering the results of Sections~III and~IV. Other practical beamformers, such as maximum-ratio transmission or MMSE beamforming, would change the effective gains and residual interference-plus-noise variances, and hence the numerical KLD, SER, and $P_{\mathrm D}$ values, though the comparative ordering across the configurations is expected to persist; a full beamformer comparison is therefore left for future work.

\begin{table*}[!t]
\centering
\caption{ Impact of imperfect CSI ($\epsilon_{\mathrm c}$)
and imperfect sensing estimation ($\epsilon_{\mathrm r}$).}
\label{tab:robustness}
\label{r28b}
\renewcommand{\arraystretch}{1.25}\scriptsize\setlength{\tabcolsep}{3pt}
\begin{tabular}{l l c c c c c c}
\hline\hline
\multirow{2}{*}{Configuration}&\multirow{2}{*}{Impairment}&$\mathrm{KLD}_{\mathrm c}^{(\mathrm{DL})}$&$\mathrm{KLD}_{\mathrm c}^{(\mathrm{UL})}$&$\mathrm{SER}^{(\mathrm{DL})}$&$\mathrm{SER}^{(\mathrm{UL})}$&$\mathrm{KLD}_{\mathrm r}$&$P_{\mathrm D}$\\
&&\multicolumn{6}{c}{\footnotesize values at $P_{\mathrm{R}}/N_0=10\,/\,20\,/\,30$ dB}\\
\hline
\multirow{5}{*}{SE-HD}
&Perfect&$25.9/30.1/30.6$&$17.3/174/1736$&$3.3/2.4/2.4{\times}10^{\text{-}3}$&$2.1{\times}10^{\text{-}2}/{<}10^{\text{-}6}/{<}10^{\text{-}6}$&$4.32/5.15/5.25$&$0.398/0.523/0.537$\\
&$\epsilon_{\mathrm c}{=}10^{\text{-}3}$&$25.9/30.0/30.6$&$17.3/173/1650$&$3.4/2.4/2.4{\times}10^{\text{-}3}$&$2.1{\times}10^{\text{-}2}/{<}10^{\text{-}6}/{<}10^{\text{-}6}$&\textemdash&\textemdash\\
&$\epsilon_{\mathrm c}{=}10^{\text{-}1}$&$22.3/25.6/26.0$&$15.2/112/308$&$5.5/4.1/4.0{\times}10^{\text{-}3}$&$3.7{\times}10^{\text{-}2}/1.9{\times}10^{\text{-}4}/1.7{\times}10^{\text{-}5}$&\textemdash&\textemdash\\
&$\epsilon_{\mathrm r}{=}10^{\text{-}3}$&\textemdash&\textemdash&\textemdash&\textemdash&$4.31/5.14/5.25$&$0.397/0.522/0.537$\\
&$\epsilon_{\mathrm r}{=}10^{\text{-}1}$&\textemdash&\textemdash&\textemdash&\textemdash&$3.94/4.70/4.80$&$0.340/0.456/0.471$\\
\hline
\multirow{5}{*}{SE-FD}
&Perfect&$33.7/38.5/39.0$&$26.0/260/2603$&$2.2/1.6/1.6{\times}10^{\text{-}4}$&$1.5{\times}10^{\text{-}3}/{<}10^{\text{-}6}/{<}10^{\text{-}6}$&$1.46/1.54/1.55$&$0.044/0.050/0.050$\\
&$\epsilon_{\mathrm c}{=}10^{\text{-}3}$&$33.7/38.4/39.0$&$26.0/259/2476$&$2.1/1.4/1.4{\times}10^{\text{-}4}$&$1.5{\times}10^{\text{-}3}/{<}10^{\text{-}6}/{<}10^{\text{-}6}$&\textemdash&\textemdash\\
&$\epsilon_{\mathrm c}{=}10^{\text{-}1}$&$29.1/32.9/33.4$&$22.8/168/463$&$3.9/2.8/2.5{\times}10^{\text{-}4}$&$3.9{\times}10^{\text{-}3}/3.5{\times}10^{\text{-}6}/{<}10^{\text{-}6}$&\textemdash&\textemdash\\
&$\epsilon_{\mathrm r}{=}10^{\text{-}3}$&\textemdash&\textemdash&\textemdash&\textemdash&$1.46/1.54/1.55$&$0.044/0.050/0.050$\\
&$\epsilon_{\mathrm r}{=}10^{\text{-}1}$&\textemdash&\textemdash&\textemdash&\textemdash&$1.33/1.41/1.42$&$0.036/0.041/0.042$\\
\hline
\multirow{5}{*}{SH-HD}
&Perfect&$100/135/140$&$17.5/175/1750$&$1.8/0.8/1.3{\times}10^{\text{-}5}$&$4.9{\times}10^{\text{-}3}/{<}10^{\text{-}6}/{<}10^{\text{-}6}$&$3.99/4.32/4.35$&$0.349/0.395/0.400$\\
&$\epsilon_{\mathrm c}{=}10^{\text{-}3}$&$99.9/135/140$&$17.5/174/1707$&$1.1/0.7/0.7{\times}10^{\text{-}5}$&$4.9{\times}10^{\text{-}3}/{<}10^{\text{-}6}/{<}10^{\text{-}6}$&\textemdash&\textemdash\\
&$\epsilon_{\mathrm c}{=}10^{\text{-}1}$&$81.5/106/109$&$15.6/130/491$&$1.8/0.8/1.5{\times}10^{\text{-}5}$&$8.1{\times}10^{\text{-}3}/{<}10^{\text{-}6}/{<}10^{\text{-}6}$&\textemdash&\textemdash\\
&$\epsilon_{\mathrm r}{=}10^{\text{-}3}$&\textemdash&\textemdash&\textemdash&\textemdash&$3.99/4.31/4.35$&$0.348/0.395/0.400$\\
&$\epsilon_{\mathrm r}{=}10^{\text{-}1}$&\textemdash&\textemdash&\textemdash&\textemdash&$3.64/3.93/3.96$&$0.297/0.340/0.344$\\
\hline
\multirow{5}{*}{SH-FD}
&Perfect&$135/178/183$&$26.2/262/2624$&${<}10^{\text{-}6}/{<}10^{\text{-}6}/{<}10^{\text{-}6}$&$1.1{\times}10^{\text{-}4}/{<}10^{\text{-}6}/{<}10^{\text{-}6}$&$3.19/3.39/3.41$&$0.234/0.262/0.265$\\
&$\epsilon_{\mathrm c}{=}10^{\text{-}3}$&$135/177/183$&$26.2/261/2559$&${<}10^{\text{-}6}/{<}10^{\text{-}6}/{<}10^{\text{-}6}$&$1.1{\times}10^{\text{-}4}/{<}10^{\text{-}6}/{<}10^{\text{-}6}$&\textemdash&\textemdash\\
&$\epsilon_{\mathrm c}{=}10^{\text{-}1}$&$111/141/145$&$23.4/195/737$&${<}10^{\text{-}6}/{<}10^{\text{-}6}/{<}10^{\text{-}6}$&$2.4{\times}10^{\text{-}4}/{<}10^{\text{-}6}/{<}10^{\text{-}6}$&\textemdash&\textemdash\\
&$\epsilon_{\mathrm r}{=}10^{\text{-}3}$&\textemdash&\textemdash&\textemdash&\textemdash&$3.18/3.38/3.41$&$0.233/0.261/0.264$\\
&$\epsilon_{\mathrm r}{=}10^{\text{-}1}$&\textemdash&\textemdash&\textemdash&\textemdash&$2.90/3.09/3.11$&$0.195/0.220/0.223$\\
\hline\hline 
\end{tabular}

\end{table*}

\section{Conclusion}
This paper developed a unified KLD-based framework for comparing deployment scenarios and duplexity paradigms in distributed CF-mMIMO ISAC systems. System models for all four configurations incorporate realistic impairments: residual SI, imperfect IC, and clutter. KLD enables direct comparison of communication and radar performance on a unified scale. It is this common scale that makes the selection rules well defined: the ordering between deployments reduces to a direct comparison of quantities measured on one scale, and the point at which FD overtakes HD becomes the crossing point of the corresponding KLD curves rather than a comparison across two metrics on different scales. Being unbounded, KLD also stays discriminative where the detection probability saturates near unity. A GLRT detection framework yielded closed-form expressions linking KLD to detection probability, with the resulting KLD--SER and KLD--detection relations validated numerically. Simulations demonstrated that FD achieves $30$--$50\%$ higher $\mathrm{KLD}_{\mathrm{c}}$ than HD when the residual IC error and SI leakage are sufficiently small, on the order of $10^{-6}$ and $10^{-3}$ or below, respectively, while HD provides a robust fallback for less stringent implementations. Although FD reduced $\mathrm{KLD}_{\mathrm{r}}$ through increased interference, detection $P_{\mathrm{D}}\approx1$ remained achievable. SH deployment raised both communication and radar performance via its larger effective aperture, though its radar gain then depended on cancellation quality, which SE deployment avoided by isolating the subsystems. These thresholds give concrete design targets for next-generation ISAC hardware. The comparative conclusions also held under imperfect CSI and sensing estimation, degrading gracefully without reordering. A complexity analysis further quantified the implementation cost of these gains, showing that SH deployment and FD operation raise the per-configuration processing burden, so that SH-FD attains the strongest performance at the greatest cost while SE or HD configurations offer a lighter alternative. Future work will extend this framework to multi-cell coordination and joint KLD-based resource allocation and waveform design.

\bibliographystyle{IEEEtran}
\bibliography{main.bib}

\end{document}